\documentclass[review,1p,times]{elsarticle}

\usepackage{lipsum}
\makeatletter
\def\ps@pprintTitle{%
	\let\@oddhead\@empty
	\let\@evenhead\@empty
	\def\@oddfoot{}%
	\let\@evenfoot\@oddfoot}
\makeatother

\usepackage{graphicx}
\usepackage{amssymb}
\graphicspath{{figures/}}
\usepackage[utf8]{inputenc}

\usepackage{amsmath}
\usepackage{accents}

\usepackage{soul,color}
\usepackage{eurosym}
\usepackage{enumitem}
\usepackage[table, xcdraw]{xcolor}
\usepackage{color}
\usepackage{tensor}
\usepackage{booktabs}
\usepackage{multirow} % for table
\usepackage{colortbl}
\definecolor{Gray}{gray}{0.9}  % Color for table cell coloring
\definecolor{DarkGreen}{rgb}{0,0.39,0}
\usepackage[T1]{fontenc}
\usepackage{lmodern,textcomp}
\usepackage{multirow}
\usepackage{soul, color}
\usepackage{bm} % math package
\usepackage{bbm} % math package
\usepackage{subcaption,siunitx,booktabs}  % tables package
\usepackage{hhline}
\usepackage{colortbl}
\definecolor{Gray}{gray}{0.9}  % Color for table cell coloring
\definecolor{DarkGreen}{rgb}{0,0.39,0}
\usepackage{eurosym}
\usepackage{color}
\usepackage{pdflscape}
\usepackage{afterpage}
\usepackage{float}

\journal{Applied Energy}

\usepackage[pagewise]{lineno}

\begin{document}

\begin{frontmatter}

\title{Data-driven Predictive Energy Optimization in a Wastewater Pumping Station}

\author[label1,label2]{Jorge Filipe}
 \author[label1]{Ricardo J. Bessa\corref{cor2}}
 \ead{ricardo.j.bessa@inesctec.pt}
\author[label1]{Marisa Reis}
\author[label3]{Rita Alves}
\author[label3]{Pedro P\'ovoa}
 \address[label1]{INESC Technology and Science (INESC TEC), Campus da FEUP, Rua Dr. Roberto Frias, 4200-465 Porto Portugal}
 \address[label2]{ Faculty of Engineering, University of Porto, Rua Dr. Roberto Frias, 4200-465 Porto Portugal}
 \address[label3]{\'Aguas do Tejo Atl\^antico, S.A., F\'abrica da \'Agua, Av. de Ceuta, 1300-254 Lisboa}
\cortext[cor2]{Corresponding author} 

\begin{abstract}
{\color{black} Urban} wastewater sector is being pushed to optimize processes in order to reduce energy consumption without compromising its quality standards. Energy costs can represent a significant share of the global operational costs (between 50\% and 60\%) in an intensive energy consumer. Pumping is the largest consumer of electrical energy in {\color{black}a} wastewater treatment plant. Thus, the optimal control of pump units can help the utilities to decrease operational costs. This work describes {\color{black}an innovative} predictive control policy for wastewater variable-frequency pumps that minimize electrical energy consumption, considering uncertainty forecasts for wastewater intake rate and information collected by sensors accessible through the {\color{black}Supervisory Control and Data Acquisition system}. The proposed control method combines statistical learning (regression and predictive models) and deep reinforcement learning ({\color{black}Proximal Policy Optimization}). {\color{black} The following main original contributions are produced: i) model-free and data-driven predictive control; ii) control philosophy focused on operating the tank with a variable wastewater {\color{black} set-point leve}l; iii) use of supervised learning to generate synthetic data for pre-training the reinforcement learning policy, without the need to physically interact with the system.} The results {\color{black}for a real case-study during 90 days} show a 16.7\% decrease in electrical energy consumption while still achieving a 97\% reduction in the number of alarms ({\color{black}tank level above 7.2 meters}) when compared with the current operating scenario ({\color{black}operating with a fixed {\color{black} set-point} level}). {\color{black} The numerical analysis showed that the proposed data-driven method is able to explore the trade-off between number of alarms and consumption minimization, offering different options to decision-makers.}
\end{abstract}

\end{frontmatter}

%%
%% Start line numbering here
%%
%\linenumbers

%% main text
\section{Introduction}
\label{S:Intro}
\subsection{Motivation}

Drivers such as climate change, population, urbanization, aging infrastructure{\color{black}s} and electricity costs are all expected to impose significant strains on urban water cycle systems management. {\color{black}The urban} water and wastewater sector is confronted by the challenge of optimizing processes in order to reduce energy consumption and reduce the emission of greenhouse gases arising from water and wastewater transport and treatment without compromising water quality standards to which they are subjected \cite{OECD2009}. It is important to highlight that energy costs represent a relevant component of water utilities operational costs and the water sector is an energy intensive consumer. For instance, the {\color{black} urban} water sector represents 1.4\% of the total electrical consumption in Portugal, including all water and wastewater infrastructures (36\% in wastewater infrastructures). In terms of total operational costs, energy consumption corresponds to around 57\% (without considering human resources). 

Additionally, the operational efficiency of wastewater services is generally lower, when compared to {\color{black}the} best practices in other industries \cite{Guerrini2015}, and it is expected that future quality standards {\color{black} will} become more restrictive and it will be necessary to improve existing treatments with the adoption of new technologies that may be intensive energy consumers. Recent advancements in smart water networks and internet-of-things are helping water and wastewater utilities to move in this direction, boosting efficiency and becoming more proactive in {\color{black}the} wastewater treatment \cite{Rasekh2016}. 

Most of the processes that occur in a wastewater treatment plant (WWTP) require electrical energy for their operation and are intensive consumers. {\color{black}Pumping} is the largest consumer of electrical energy in WWTP \cite{Kebir2014}. Therefore, the control of pump units in WWTP can help utilities to decrease operational cost{\color{black}s} (e.g., electricity cost{\color{black}s}) providing increased energy savings and environmental performance. The predictive control of pumping systems is the main objective of the present work. 

\subsection{Related Work and Contributions}

Kalaiselvan et al. conducted a literature review of energy efficiency actions for pumping systems, grouped by component design, selection and dimensioning, control and adjustment of variable-frequency pump units \cite{Kalaiselvan2016}. The category ``control and adjustment'' was divided in the following sub-categories: (a) variable-frequency drive control; (b) load shifting; (c) process optimization.

Load shifting has been an active area of research, mainly focused on the improvement {\color{black}of} the mathematical tractability and performance of mixed integer-nonlinear programming for scheduling the operation of pumping stations in water distribution systems, over a future period (e.g., day-ahead), {\color{black}in order} to minimize electrical energy costs subject to constraints that account for the distribution system hydraulics {\color{black}modelled with the hydraulic simulation code KYPIPE} \cite{Brion1991}. {\color{black}For the same problem, Menke et al. proposed linearized (approximated) optimization models solved using a branch and bound method that generate optimized pump schedules with bounded optimality gaps.}~\cite{Menke2016a}. {\color{black}Another example is the model predictive control (MPC) with binary integer programming optimization for load shifting in a water pumping system} \cite{Jacobusvan2011}. 

Alternative approaches to the classical mathematical programming are meta-heuristics, such as: {\color{black}multi-objective evolutionary algorithm that minimizes the cost of pumping and maximizes the minimum stop time} \cite{Lopez-Ibaez2005}; {\color{black}genetic algorithms (GA) for optimizing the pumping schedules of water supply systems, combined with EPANET hydraulic solver to assess the feasibility of potential solutions}~\cite{Kernan2017}; {\color{black} \cite{Torregrossa2019} compared GA, simulated annealing and particle swarm optimization (PSO) for dynamically managing the pump activation heights that minimize  energy costs over a given time horizon, considering a physical modelling of the system that includes cavitation and overflow features.}

This problem is gaining new attention with demand-side management programs and opportunities created by the synergy between smart electric and water distribution networks \cite{Palensky2011}. This new paradigm requires tractable convex mixed-integer non-linear programming (MINLP) problems, robust to highly dynamic electricity tariffs~\cite{Bonvin2017}, or the use of different variants of dynamic programming for solving the optimization problem~\cite{Zhuan2013}. Furthermore, opportunities like the participation in ancillary services (i.e., short term operational reserve, firm frequency response, frequency control by demand management) will emerge for flexible pumping systems \cite{Menke2016}.    

All the aforementioned works share the following characteristics: i) are focused {\color{black}o}n operational scheduling of water distribution systems; ii) do not cover real-time control of pumping units (i.e., continuous optimization) and are designed for fixed-speed motored pumps (i.e., integer variables); iii) the formulations rely {\color{black}o}n approximations of the hydraulic model. Moreover, in WWTP, ``long-term'' flexibility (i.e., load shifting) is lower in comparison to water distribution. Therefore, most of the potential energy savings are a result of ``short-term'' flexibility, i.e., energy optimization close to real-time of variable-frequency pumps. For instance, the MPC, proposed by van Staden et al. in \cite{Jacobusvan2011}, can only be applicable in practice under certain conditions: i) constant or known water inflow rate; ii) large water reservoirs that allow pump scheduling in a 24 hours window (e.g., considering different electricity tariffs); iii) binary (on/off) control of pumps.

The present paper fits in the ``process optimization'' category and proposes an innovative data-driven energy optimization strategy of WWTP variable-frequency pumps. It combines statistical learning (regression and predictive models) and artificial intelligence (AI) techniques. The main objective is to design predictive pump control policies that minimize electrical energy consumption, considering uncertainty forecasts for wastewater intake and information collected by sensors that are typically installed in WWTP and accessible through the SCADA system. Constraints related to the desired output (i.e., wastewater output flow/head) are also included. Furthermore, this control approach does not require an explicit (or mathematical) model of the process, since the system dynamics are learned from the data. 

{\color{black}Modelling the physical system with supervised learning has been explored by other authors in the MPC framework and for other problems. Afram et al. proposed the use of artificial neural networks for modelling Heating, Ventilating and Air Conditioning (HVAC) systems, which are integrated in an optimization problem solved with different approaches, such as GA, PSO, interior-point method, branch and bound \cite{Afram2017}. Jain et al. propose the use of online Gaussian processes for real-time closed-loop finite horizon receding horizon control, which takes advantage from optimal experiment design when limited data are available for training \cite {Jain2018}. This method was applied to energy management during building load curtailment for economic demand response. Linear models were also explored by other authors, such as regression trees (i.e., dataset partition until having leaves where linear parametric models can be applied) and random forests for receding horizon control applied to building energy management~\cite{Smarra2018a}. The same authors proposed in \cite{Smarra2018} a state-space switched affine dynamical model that outperforms static models constructed with regression trees or random forests. In contrast to these modelling methods for MPC, the data-driven modelling methodology proposed in the present paper is oriented to the RL paradigm, which does not require a close-form expression like MPC, and is developed specifically for wastewater pumping stations.}

For this problem, the industry{\color{black}'s} state-of-the-art is to turn on/off pumps {\color{black} or adjust operating point of variable-frequency pumps} according to a {\color{black}fixed} level-based control system. For instance, this patent \cite{nybo2014method} describes a method for operating a pumping system of a WWTP where the pump starts operating if a level of a wastewater in a tank exceeds a first level, and the pump stops pumping if the level of the wastewater in the tank drops below a second level. However, control algorithms based on soft computing are earning attention and being explored for WWTP and water distribution systems, mainly in cases were the physical (or mathematical) model is not available or is too complex to be integrated in a classical controller. {\color{black} It is important to underline that data-driven techniques, such as fuzzy logic, clustering, neural networks, support vector machines, have been applied to the wastewater sector for many years and one {\color{black}of the} driving forces was to increase the performance of processes and optimize the usage of resources \cite{Corominas2018}. Moreover, these techniques can also be used to derive user-friendly performance index for pump energy efficiency and identify potential failures at an early stage \cite{Torregrossa2017}.} 

Fiter et al. described a fuzzy logic controller to regulate the aeration in the bioreactor of a WTTP \cite{Fiter2005}. The controller integrated information from two signals, dissolved oxygen and oxidation-reduction potential values, to minimize the electrical energy consumption. Reinforcement learning (RL) was also applied to optimize different processes in WWTP. Syafiie et al. proposed a model-free control approach for advanced oxidation processes (or Fenton process){\color{black},} since according to the authors it is extremely difficult do develop a precise mathematical model and the system is {\color{black}subject} to several uncertainties and time-evolving characteristics \cite{Syafiiea2011}. As an alternative to proportional–integral–derivative (PID) controllers, Hern\'andez-del-Olmo et al. explored RL for oxygen control in the N-ammonia removal process, {\color{black}whose} main objective was to minimize WTTP operational cost{\color{black}s} (including energy costs) \cite{del-Olmo2012a}. Asadi et al. optimized water quality and energy consumption of the aeration process by combining boosting trees for feature selection and different machine learning algorithms (e.g., artificial neural networks, random forests) for modeling the relationship between input, controllable and output variables \cite{Asadi2016}. {\color{black}Han et al. proposed a multi-objective control strategy that satisfies the effluent quality and reduces the energy consumption of the wastewater treatment process \cite{Han2018}. The control method combines multi-objective PSO and uses a fuzzy neural network controller to trace the obtained set-points for dissolved oxygen and nitrate (fully detailed in \cite{Han2017}).}

More related with the work of the present paper, Kebir et al. described a rule-based method to control WWTP pumps according to the measured wastewater tank level and minimize electrical energy consumption by using a fuzzy logic controller that work{\color{black}s} as follows: on rainy day{\color{black}s}, pumps react faster and frequency is increased quickly to avoid flooding; on dry day{\color{black}s,} pumps can react more slowly and frequency is decreased softly to prevent {\color{black}the} draining of the tank \cite{Kebir2014}. 

Data mining algorithms were also explored to optimize the pump operation, including the modeling of input and output variables. Wei and Kusiak applied static multi-layer perceptron and dynamic neural networks to forecast the influent flow in WWTP \cite{Wei2015}. Zhang and Kusiak tested seven data mining algorithms based in 5-min and 30-min data to construct a pump energy consumption model and a water flow rate (after the pumps) model of the preliminary treatment process of a WWTP \cite{Zhang2011}. These two data-driven models were incorporated in different formulations of an optimization problem to generate optimal pump schedules: (a) MINLP problem to reduce energy consumption solved with PSO  \cite{Zhang2012} or with greedy electromagnetism-like algorithm \cite{Zeng2016}; (b) bi-objective optimization solved with artificial immune network algorithm to minimize the energy consumption and maximize the pumped wastewater flow rate \cite{Zhang2016}. These optimization models can be enhanced with a discrete-state Markov process for modeling the maintenance decisions \cite{Zhang2015}.

Considering the revised literature, the present paper produces the following original contributions. Applies a model-free and data-driven control approach based in RL, in contrast to the use of meta-heuristics \cite{Zhang2012,Zeng2016,Zhang2016} or fuzzy logic control \cite{Kebir2014}. The control philosophy is focused {\color{black}o}n operating the tank with a variable {\color{black} set-point} wastewater level, instead of controlling the frequency increase/decrease rate like in \cite{Kebir2014}. A real-world implementation of the RL control method is made possible by applying data-mining algorithms to construct models from data and generate synthetic data for pre-training the RL algorithm, without the need to physically interact with the system. This also represents an original contribution compared to other control problems like \cite{del-Olmo2012a,Syafiiea2011}. {\color{black} In order to address this problem and in a different context for building heating systems, an interesting and different approach is proposed by Costanzo et al. \cite{Costanzo2016}, and inspired by the work in \cite{Lampe2014}, which consists in using a neural network to create virtual data that are used together with experimental data to approximate the Q-function (state-action value function) of fitted Q-learning and reduce the amount of system interaction required to learn a ``good'' policy.}

{\color{black}Moreover, in contrast to \cite{Wei2015}, where forecasts are explicitly included in the optimization model, the vector of system states from RL is modified to include probabilistic forecasts of wastewater intake and implement a predictive control of the pumping system. A similar approach can be found in \cite{Costanzo2016} for fitted Q-iteration RL algorithm where exogenous variables (e.g., ambient temperature, irradiance) are replaced by forecasted values}. 

Finally, the proposed method is applied to a real case-study and avoids the use of simplifications or simulation models for the physical system.

\subsection{Structure of the Paper}

The remaining of the paper is organized as follows: section \ref{sec:process_description} describes the wastewater treatment process and a real-world pumping station; section \ref{sec:framework} presents the data-driven control framework and its different modules; section \ref{sec:control_method} presents the environment emulation method and predictive control method; the numerical results are discussed in section \ref{sec:results} and {\color{black}the} main conclusions {\color{black}are} provided in section \ref{sec:conclusions}.

\section{Wastewater Treatment Process Overview}\label{sec:process_description}

A wastewater treatment system is composed by a drainage system, a WWTP and the discharge infrastructure. Traditionally, a WWTP is divided into mechanical, physical, chemical and biological treatment{\color{black}s}, which has been utilized with many different combinations and organized in different treatment phases \cite{Yoo2001}: pre-treatment, primary treatment, secondary treatment (biological), tertiary treatment, disinfection, sludge treatment and odor control.

\begin{figure}[hbt]
	\centerline{\includegraphics[scale=0.40]{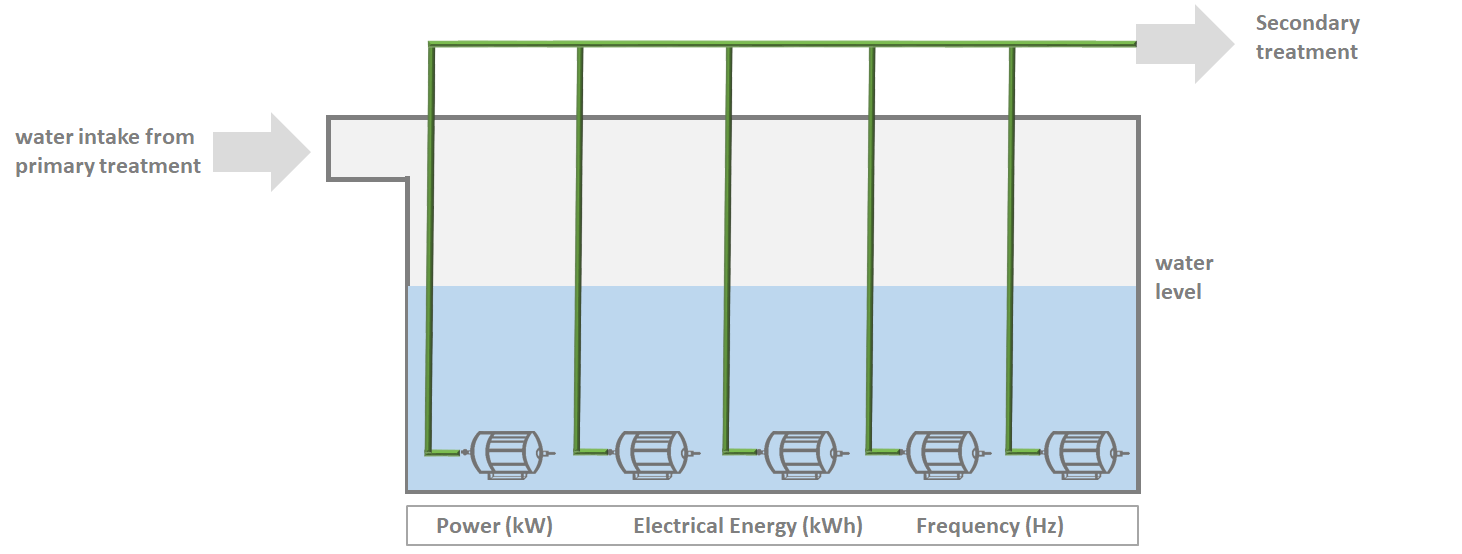}}
	\caption{Intermediate wastewater pumping station with five variable-frequency pumping units of 110 kW.}
	\label{fig:water_tank}
\end{figure}

This work is focused {\color{black}on} the control of wastewater pumping stations in a WWTP, F\'abrica da \'Agua de Alc\^{a}ntara, Portugal. The case-study corresponds to the intermediate wastewater pumping station (WWPS) between primary and secondary treatment, which is depicted in Figure \ref{fig:water_tank}. The primary treatment involves the separation of {\color{black}suspended} solid matter from the wastewater using {\color{black} lamellar} primary clarifiers {\color{black}with chemical addition}. This solid matter {\color{black}named primary sludge} is pumped out of the tanks for further treatment and the remaining water is {\color{black} conducted to the feed channels of the biological treatment, located at the top of the biofiltration cells, by the intermediate pumping station. {\color{black}Secondary treatment is applied to degrade the remain dissolved organic and inorganic matter by biological aerobic filter. The WWPS} is equipped with five submersible electric pump groups controlled by variable frequency drivers and the frequency and active power measurements are collected for each pump. The primary treatment effluent flow (wastewater intake flow) is measured in 5 electromagnetic located in the primary treatment pipes. An electromagnetic flow meter installed in the compression pipeline measures the total inflow to the biological treatment (WWPS effluent flow). The flow variation can range from 500 l/s to 4000 l/s.} 

{\color{black}The WWPS dataset (pump active power, frequency, tank level, wastewater intake and outflow rate) comprises 470,962 observations in the period between September 2013 and June 2017 with non-constant sampling frequency. About 68.9\% of the observations are sampled with 2 minutes interval, followed by 3 minutes sampling with 16.1\%, and 8.5\% of observations above 5 minutes.}

{\color{black} The current pump control policy of the WWPS consists in operating with a reduced variation level of $\pm10\%$ around 6 meters, considering: i) a bottom limit of 2.5 meters for pump safety; ii) a first alarm at 7.2 meters that initiates a preventive action that re-directs a share of the incoming wastewater directly to the output of the facility, skipping some treatment phases; iii) a second alarm occurs at 8 meters when the tank reaches its maximum capacity and overflow occurs. For this purpose, a level probe, which measures the tank level in meters, is installed.} 

{\color{black} Presently, the automatic pump operation is carried out according to the respective level measurement, level set-point value (i.e., 6 meters) and pump sequence selection. Due to the deviation between the value of the measured level and the level set-point, a frequency control signal is sent to the pump(s).}

{\color{black} If the level goes above 6 meters:
\begin{enumerate}
\item The first pump of the sequence is started and its operating frequency is set according to the set-point level entered.
\item If the first pump reaches the rated frequency (50 Hz), a second pump is started. Both pumps will operate at the same frequency.
\item If the first two pumps have reached rated speed, a third pump is started. The pump that started first will run at 50 Hz, and the other two (2nd and 3rd) will adjust their frequency according to the set-point level entered.
\item If the first three pumps start running at 50 Hz, the fourth pump will start. In this situation, the first two will work at a fixed speed of 50 Hz.
\item The start-up sequence will be carried out by the criterion of the pump being stopped for the longest time.
\item The fifth pump associated with the soft starter and therefore working at fixed speed, will only start operating when three or four pumps are required to operate. 
\end{enumerate}}

{\color{black} If the level drops below 6 meters:
\begin{enumerate}
\item The stop sequence respects the start sequence.
\item The last two pumps to start operating with frequency driver will slow down their speed. If the minimum operating frequency is reached, a stop command is given to the pump that started in the first place.
\item When the last two pumps with frequency drivers reach the minimum speed, the pump running longer stops.
\end{enumerate}
}

Finally, it is important to stress that this case-study consists of a single pumping station, but the wastewater treatment system has a total of 13 analogous processes, and the company a total of 292 stations. Therefore, in order to replicate the proposed control policy, it is only necessary to have variable-frequency pump units and the variable collected by the SCADA system (see next section).

\section{Data-driven Predictive Control Framework}
\label{sec:framework}

The proposed data-driven control framework aims at optimizing the energy consumption of WWPS by optimally defining the operating set-point for each variable-frequency pump unit. Given the layout of the WWPS, described in section~\ref{sec:process_description}, electrical energy gains can be achieved by operating with a higher wastewater level, in order to reduce the relative height between the wastewater tank level and the secondary treatment tank. However, operating with a higher level also increases the risk of wastewater overflow, due to the uncontrollable and volatile rate of the wastewater intake from the primary treatment.

Presently, wastewater pumping stations are operated with fixed-level control rules. This sub-optimal solution has the same buffer (difference between the maximum height of the tank and the 7.2 meters alarm), independently of the season. In dry seasons there is a lower wastewater intake rate (WWIR), therefore the station could be operated with a higher level without impacting the safety of the operation. On the other hand, during wet seasons, which have a much higher and volatile WWIR, {\color{black}it could dynamically reduce} the wastewater level to accommodate extreme WWIR.

The proposed predictive control has the ability to anticipate the incoming WWIR and to adjust the reservoir buffer accordingly. For this, the algorithm relies {\color{black}o}n two main functions: i) WWIR forecasting {\color{black}described in section~\ref{sec:forecasting}}{\color{black},} and ii) AI control based on RL, {\color{black}described in sections~\ref{sec:emulation}-\ref{sec:implementation_stage} and~\ref{sec:control_method}}. Probabilistic forecasts of the WWIR are generated and used as one of the inputs of the RL control algorithm{\color{black},} in order to provide {\color{black} the control strategy with information about WWIR forecast uncertainty.} 

{\color{black}\subsection{General Description of the Methodology}\label{sec:framework}}

\begin{figure}[H]
	\centerline{\includegraphics[scale=0.5]{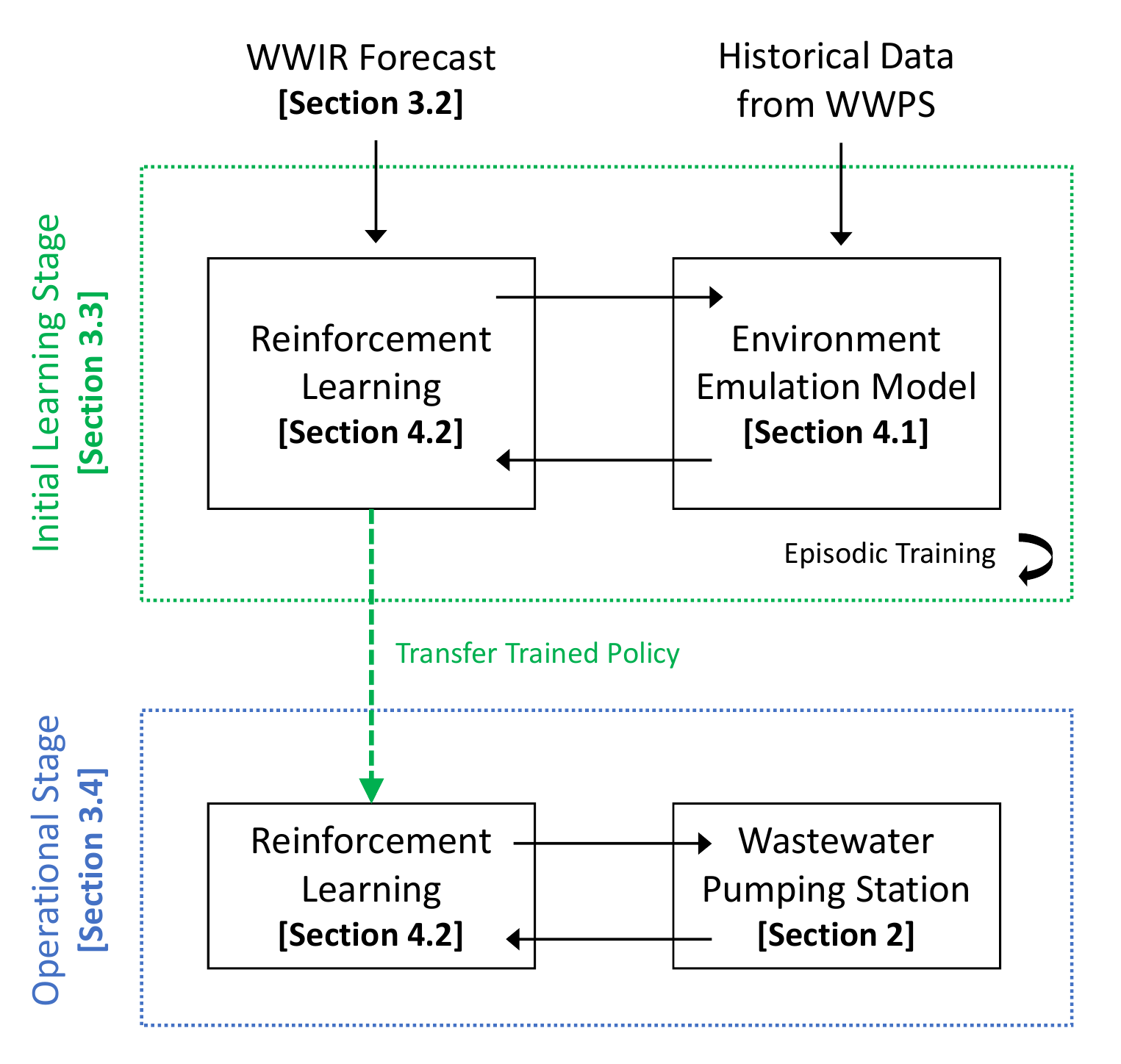}}
	\caption{{\color{black}Overall scheme of the predictive control methodology for wastewater pumping stations.}}
	\label{fig:single_framework}
\end{figure}

{\color{black} Figure \ref{fig:single_framework} depicts the overall scheme of the predictive control methodology for WWPS. The building blocks of the control philosophy are the WWIR forecasting model, RL algorithm and environment emulation model.} 

The AI control relies on RL concepts. RL is a machine learning method which contrasts heavily with more traditional AI classes, such as supervised learning (SL). In SL, the fitting of a model is made through instructive feedback, so there is a tangible path to improve the model (loss function). In RL, the learning process is made by evaluative feedback, which translates in knowing how well you {\color{black}have} achieved your goal (reward). For each learning instant, the RL algorithm uses as input a set of variables that are able to provide a snapshot of the environment and these values are used to sample an action. This action will result in a state transition that will produce a reward. Through several interactions with an environment, the control policy will learn the optimal action or course of actions that maximize the expected reward.

\begin{figure}[H]
	\centerline{\includegraphics[width=0.7\textwidth]{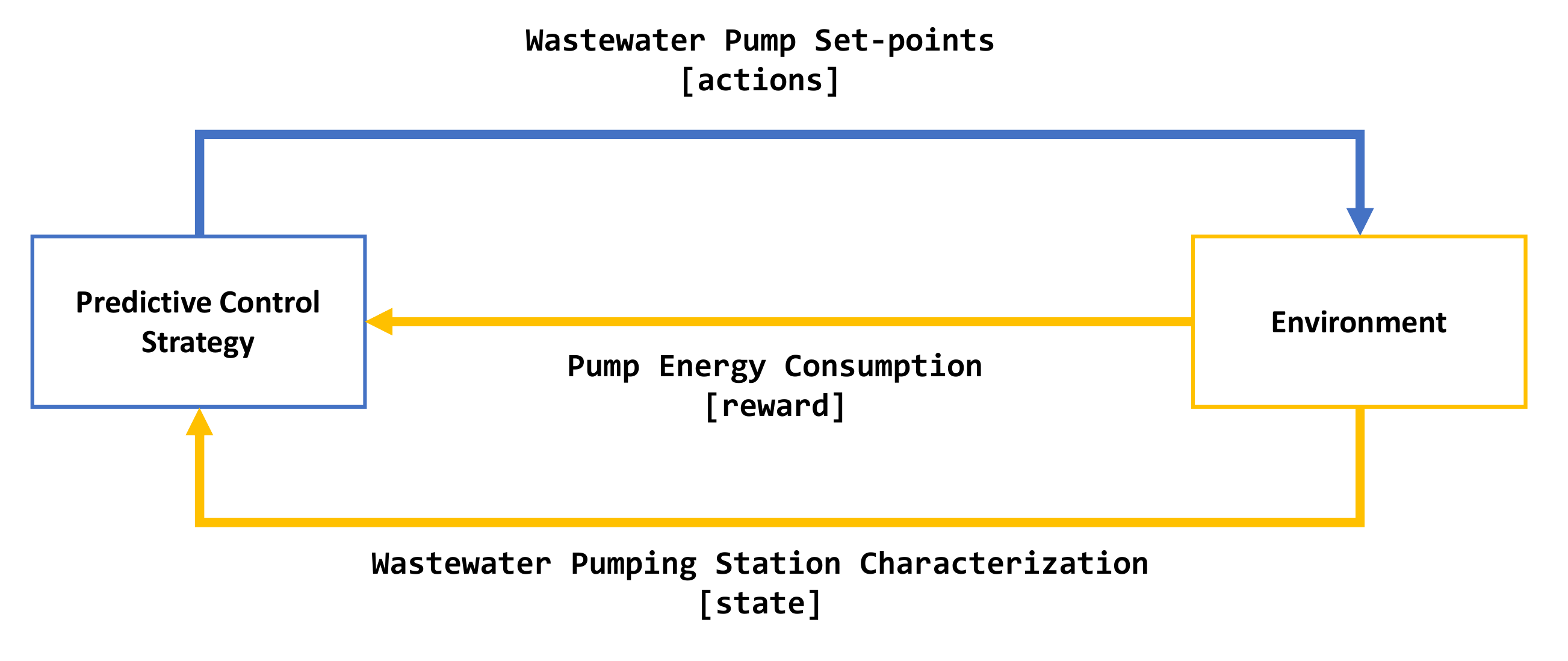}}
	\caption{{\color{black}Reinforcement learning framework applied to control wastewater pumping station}.}
	\label{fig:esquema_RL_basic}
\end{figure}

{\color{black} Figure~\ref{fig:esquema_RL_basic} illustrates the learning process adapted for the WWPS (i.e., environment). From the environment{\color{black},} a state vector is constructed using data (e.g. tank or reservoir level, WWIR forecasts, pumps online and current operational set-point, etc.) acquired by a set of sensors and is fed to the predictive control strategy. With this information, the algorithm is able to select actions (power set-point for each pump unit), which are applied in {\color{black}the} environment. The reward that provides feedback and allows the algorithm to learn, is a reflection of the pumps' electrical energy consumption.}

{\color{black}RL requires thousands of interactions with the environment, exploring different actions until developing an efficient control strategy.} When dealing with physical systems composed by expensive mechanical components and in continuous operation, it is impossible to directly implement a control algorithm that relies on thousands of interactions with the system to learn the optimal control policy. 

{\color{black} Therefore, as depicted in Figure~\ref{fig:single_framework}, a two-stage framework is proposed: 1) the model is trained on a{\color{black}n} emulated environment{\color{black},} constructed with SL, i.e. \textit{initial learning stage}; 2) the trained policy is transferred to an operational setup and applied to the physical wastewater pumping station, i.e. \textit{operational stage}.

In the initial learning stage, using historical records acquired by the SCADA system, two statistical learning models, together, are able to translate pumping power into wastewater outflow rate and, consequently, estimate the variation of the wastewater tank level. The emulated system can sustain every possible action of the RL methodology, hence allowing the algorithm to explore every possible range of actions without adding stress to the physical system.

During the operational stage, the RL model is applied to the physical system and is able to fine tune the pre-trained model by continuous learning{\color{black},} in order to mitigate mismatches between the emulated and physical systems.

These building blocks are described in the following sections.}

{\color{black}\subsection{Forecasting Wastewater Intake Rate}\label{sec:forecasting}}

{\color{black}Statistical learning methods and feature engineering (i.e., {\color{black}the} creation of additional features from the raw dataset) are used to generate point (i.e., expected value) and probabilistic multi-step ahead forecasts for the WWIR.}

\subsubsection{Data Pre-processing and Feature Engineering}\label{sec:feature_eng}

As mentioned in section~\ref{sec:process_description}, an important characteristic of the time series{\color{black}, and which have} impact in the feature creation{\color{black},} is that measurements are collected with irregular sampling frequency. {\color{black}This} is particularly challenging for including lagged variables (i.e., past values of the time series) in the forecasting model, according to an autoregressive framework. To overcome this limitation and enable the inclusion of lagged variables, a re-sampling is performed over a uniform time interval. In this case, it was re-sampled to the most common observed time frequency, which was 2 minutes. 

{\color{black}The autocorrelation analysis provides the correlation as a function of the lags when comparing the original series with the lagged version of itself and showed a daily seasonality. Based on this information, lagged features were added to the model, $W_{t-1}, ..., W_{t-l}$ where $l$ is equal to 8 as determined in Appendix A, as well as the WWIR from the same timestamp of the previous day,  denoted as ${}^{24H}_{}W^{}_{t}$.} 

{\color{black}The average WWIR strongly depends on the period of the year and period of the day. Higher variability was observed in the morning period. For the monthly analysis, higher values are observed for winter months, e.g. in 2014, January and February showed an average 9816 $m^3/h$ and 10832 $m^3/h$ intake, respectively.} The monthly variation is mainly due to higher precipitation levels. This analysis suggests that calendar variables, like the hour of the day, month and weekday{\color{black},} should be included {\color{black}in} the forecasting model. 

In addition to the classical lagged and calendar variables, from the raw variables dataset it is possible to derive additional features. Considering the WWIR time series represented by $W = \{W_1, ..., W_n\}$ and its timestamp represented by $\{t_1,...,t_n\}$, the following new features were calculated as follows:
\begin{itemize}
{\color{black}
\item Change over Time (CoT):
\begin{equation}
c_{m} = (W_{m-1}-W_{m})/(t_{m-1}-t_m)
\end{equation}
\item Growth or Decay (GoD): 
\begin{equation}
(W_{m-1}-W_{m})/W_m
\end{equation}}
\end{itemize}

The features considered for the forecasting model are (a detailed analysis of the variable importance can be found in Appendix A): i) calendar variables such as hour, weekday and month; ii) lagged variables close to launch time $t$ \textendash $W_{t-1}, ..., W_{t-l}$ \textendash and 24 hour lag \textendash ${}^{24H}_{}W^{}_{t}$, iii) difference series translating change \textendash $CoT_{t-1}$ \textendash and slope growth or decay \textendash $GoD_{t-1}$.
Note that except for the calendar variables, all features are {\color{black}built} backward-looking, in the sense that each point of the time series only depend on past values. 

\subsubsection{Forecasting Model}

The objective of the forecasting model is to obtain a model that approximates an unknown regression function $W=f(\textbf{x})$, where $\bf x$ is the set of features described in the previous section. In order to produce multi-step ahead forecasts, a model is fitted for each lead-time $t+k$ as follows:

\begin{eqnarray}
\begin{matrix}
\hat{q}_{t+1} = f_1 (W_{t-1}, ..., W_{t-l}, hour, wday, month, CoT_{t-1}, GoD_{t-1}, {}^{24H}_{}W^{}_{t+1}) \\
\hat{q}_{t+2} = f_1 (W_{t-1}, ..., W_{t-l}, hour, wday, month, CoT_{t-1}, GoD_{t-1}, {}^{24H}_{}W^{}_{t+2}) \\
\vdots \\
\hat{q}_{t+k} = f_k (W_{t-1}, ..., W_{t-l}, hour, wday, month, CoT_{t-1}, GoD_{t-1}, {}^{24H}_{}W^{}_{t+k})
\end{matrix}
\end{eqnarray}
where $l$ is the number of lags, $t+k$ is the lead-time horizon and $\hat{q}_{t+k} \equiv \hat{q}^{(\alpha_i)}_{t+k}$ denotes the forecast for the quantile with nominal proportion $\alpha_i$ issued at time $t$ for forecast time $t + k$. In order to produce these forecasts, $f$ is fitted for each step ahead with information available at time $t$ for the $k$ horizons.

In this work, two different statistical learning algorithms were considered for modeling function $f$:
\begin{itemize}
\item Linear Quantile Regression (LQR) \cite{QR1978}, which is a linear model analogous to multi-linear regression but with the possibility to adjust a specific model to generate a conditional quantile estimation. This model is available in the \textsc{statsmodels} library\cite{statsmodels}. 
\item Gradient Boosting {\color{black}}Trees (GBT) \cite{Friedman2001GBT}, which is an ensemble of regression trees as base learners and presents a high flexibility in accepting different types of loss functions. In this work, the quantile loss function is used to generate probabilistic forecasts. The model is available in the \textsc{scikit-learn} library \cite{Pedregosa2011}.
\end{itemize}

{\color{black}The numerical results of the forecasting skill are presented in Appendix A. Since the GBT model presents the best performance in all metrics, the probabilistic forecasts generated from this model are used in predictive control strategy described in the following sections.}

\subsection{Initial Learning Stage}
\label{sec:simulation_stage}

\begin{figure}[H]
	\centerline{\includegraphics[scale=0.45]{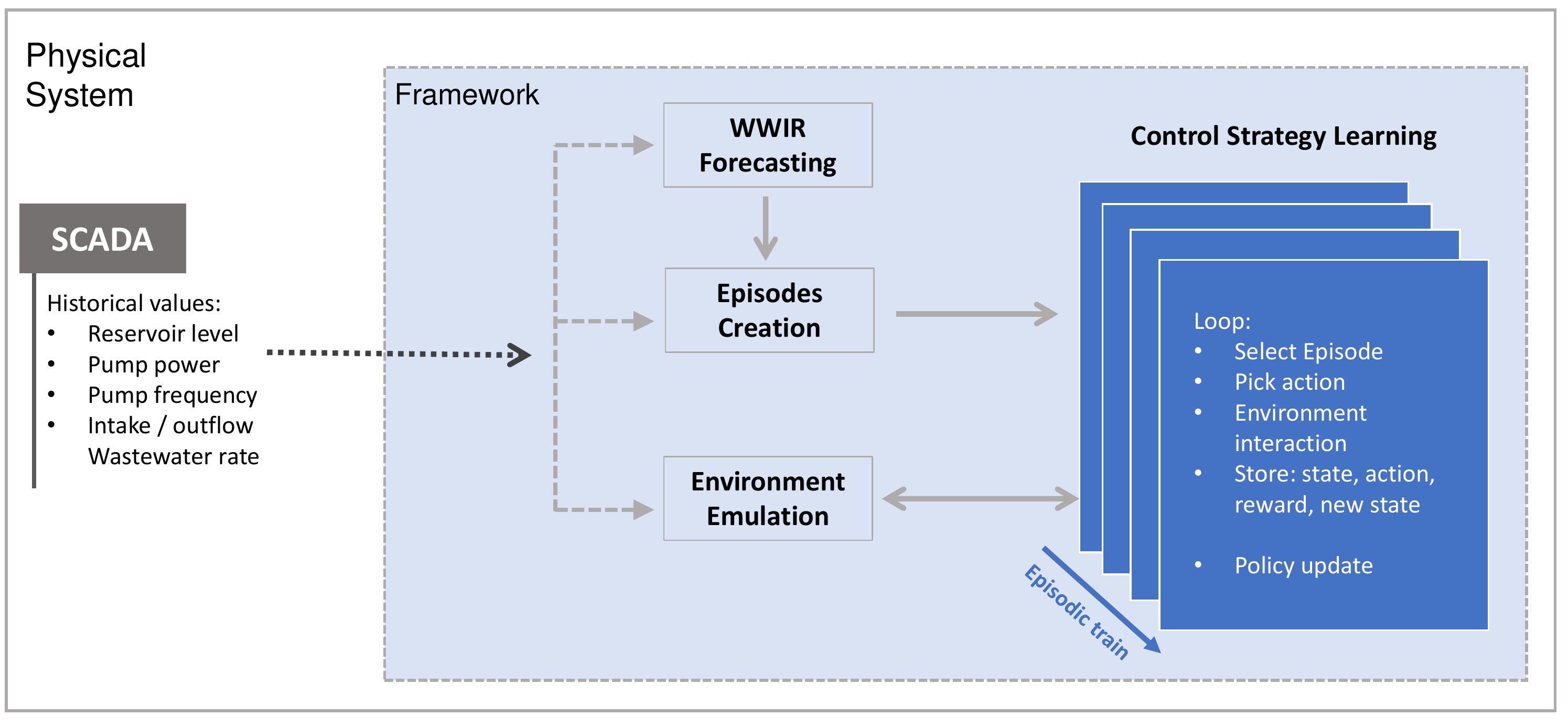}}
	\caption{{\color{black}Initial learning stage of the data-driven control method where SCADA historical data is used to: generate wastewater intake rate forecasts; construct data-driven models that emulate the physical environment; create episodes combining historical data and forecasts.}}
	\label{fig:simulation_framework}
\end{figure}

This first stage is responsible for conducting an initial learning of the control policy. Historical data is collected from the SCADA system of the WWPS, containing information about the pumps (active power and frequency) and the wastewater tank (reservoir level, WWIR and outflow rate). This data is used in three modules: WWIR forecasting, episodes creation and environment emulation, as depicted in Figure~\ref{fig:simulation_framework}.

\begin{itemize}
\item The forecasting module uses the historical time series of WWIR in order to produce probabilistic forecasts (see section \ref{sec:forecasting}). 
\item The WWPS operation is continuous. However, in order to maximize the historical information used to train the control algorithm, the complete time series was divided into a set of episodes with unequal length. Each step represents {\color{black}a} 2-minute discrete interval of the WWPS and has the following information: pumps online, current pump set-point, WWIR, original wastewater level and WWIR forecast for 20 steps ahead (40 minutes). At the start of each episode, the initial wastewater level is randomized in order to increase the diversity of the learning dataset. These episodes are then sampled randomly and used in the learning process of the RL control.
\item The environment emulation module applies statistical learning algorithms to the historical data in order to emulate the  physical processes: i) relation between pumping power and outflow rate and ii) modeling of the reservoir level deviation as a function of the WWIR and outflow rates. Both methods are detailed in section~\ref{sec:emulation}.

\end{itemize}

To summarize, from the historical data collected via SCADA system{\color{black},} the following information is created: i) WWIR probabilistic forecasts; ii) data-driven models that emulate the physical environment; iii) episodes combining historical data and forecasts.

The learning process uses this information in the following way:
\begin{enumerate}
\item Select an episode
\item Randomize the initial wastewater level
\item For each step of the episode:
\begin{enumerate}
\item Collect the state characterization \\[-25pt]
\item Sample an action from the control policy \\[-25pt]
\item Apply the action to the emulated environment  \\[-25pt]
\item Observe the state transition \\[-25pt]
\item Collect the reward \\[-25pt]
\item Store the vector [state, action, reward, new state] \\[-25pt]
\end{enumerate}
\item Update the  control policy (according to section~\ref{sec:control_policies})
\end{enumerate}

After thousands of episodes, the control policy will learn which action (or set of actions), for a given state, results in the {\color{black}highest} expected reward in the long term. 

\subsection{Operational Stage}
\label{sec:implementation_stage}

After the initial learning stage, the control policy is optimized and ready to be integrated with the physical system. However, due to inconsistencies between the  emulated and physical system, some interactions with the real environment are necessary to calibrate the policy. 

\begin{figure}[H]
	\centerline{\includegraphics[scale=0.45]{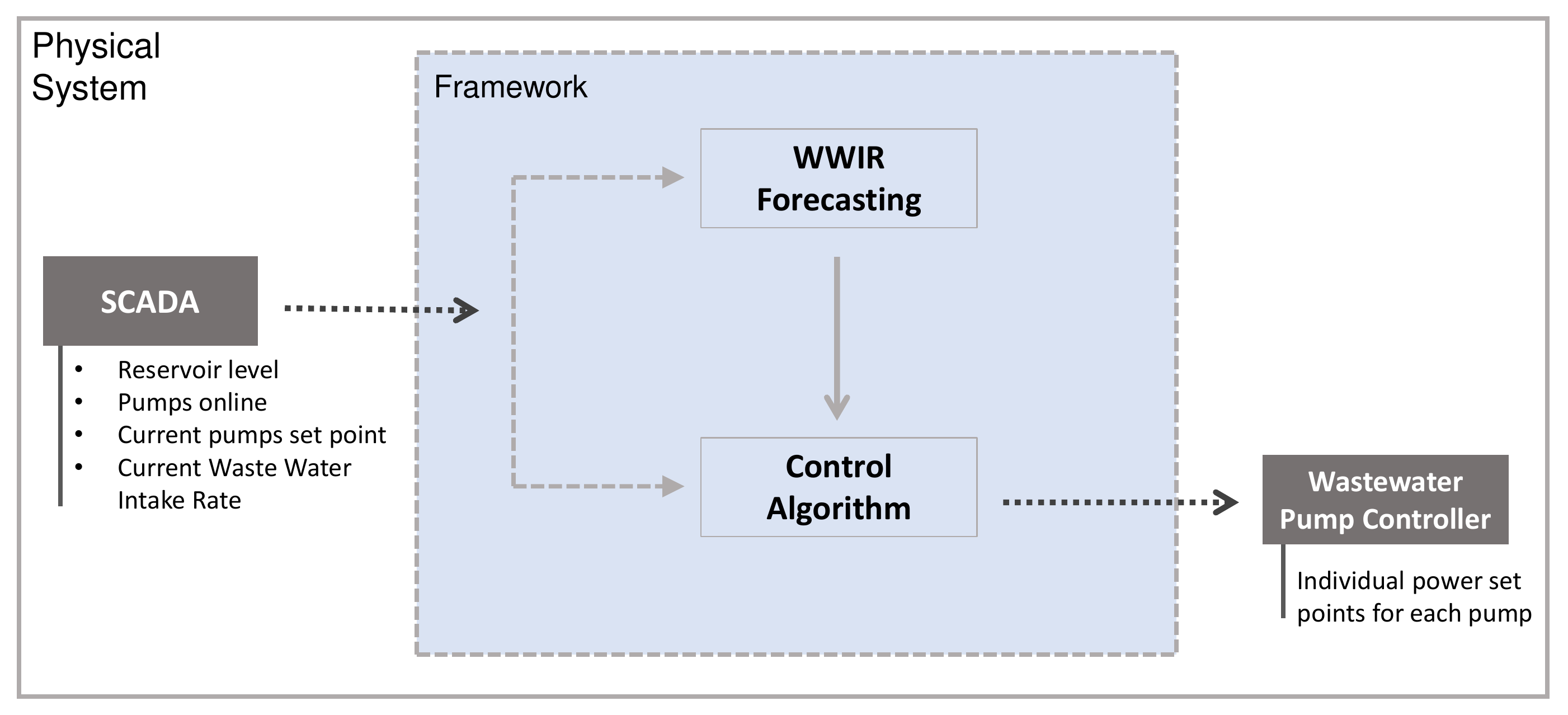}}
	\caption{Operational stage of the data-driven control framework {\color{black}where wastewater intake rate forecasts are generated and the reinforcement learning control policy is applied to the physical environment.}}
	\label{fig:implementation_framework}
\end{figure}

During the operational stage, the overall process, as depicted in Figure~\ref{fig:implementation_framework}, is the following:
\begin{enumerate}
\item Collect the state characterization from the SCADA \\[-25pt]
\item Generate the WWIR probabilistic forecasts \\[-25pt]
\item Sample an action from the control policy \\[-25pt]
\item Apply the action to the physical environment  \\[-25pt]
\item Observe the state transition \\[-25pt]
\item Measure the power consumption (reward) \\[-25pt]
\item Store the vector [state, action, reward, new state] \\[-25pt]
\item Update the  control policy (according to Section~\ref{sec:control_policies})
\end{enumerate}

\section{Process Modeling and Predictive Control}\label{sec:control_method}

\subsection{Environment Emulation Model}
\label{sec:emulation}

As mentioned before, the RL methods require thousands of interactions with the environment in order to find the action{\color{black},} or course of actions{\color{black},} that maximize the expected reward. Since it is impractical, from an implementation point of view, to learn and interact from scratch directly with the physical system, it is necessary to emulate the environment. This emulation relies {\color{black}o}n the two modules described in sections \ref{sec:wwout} (wastewater outflow rate) and \ref{sec:reser_level} (wastewater level of the tank). 

\begin{figure}[H]
	\centerline{\includegraphics[scale=0.45]{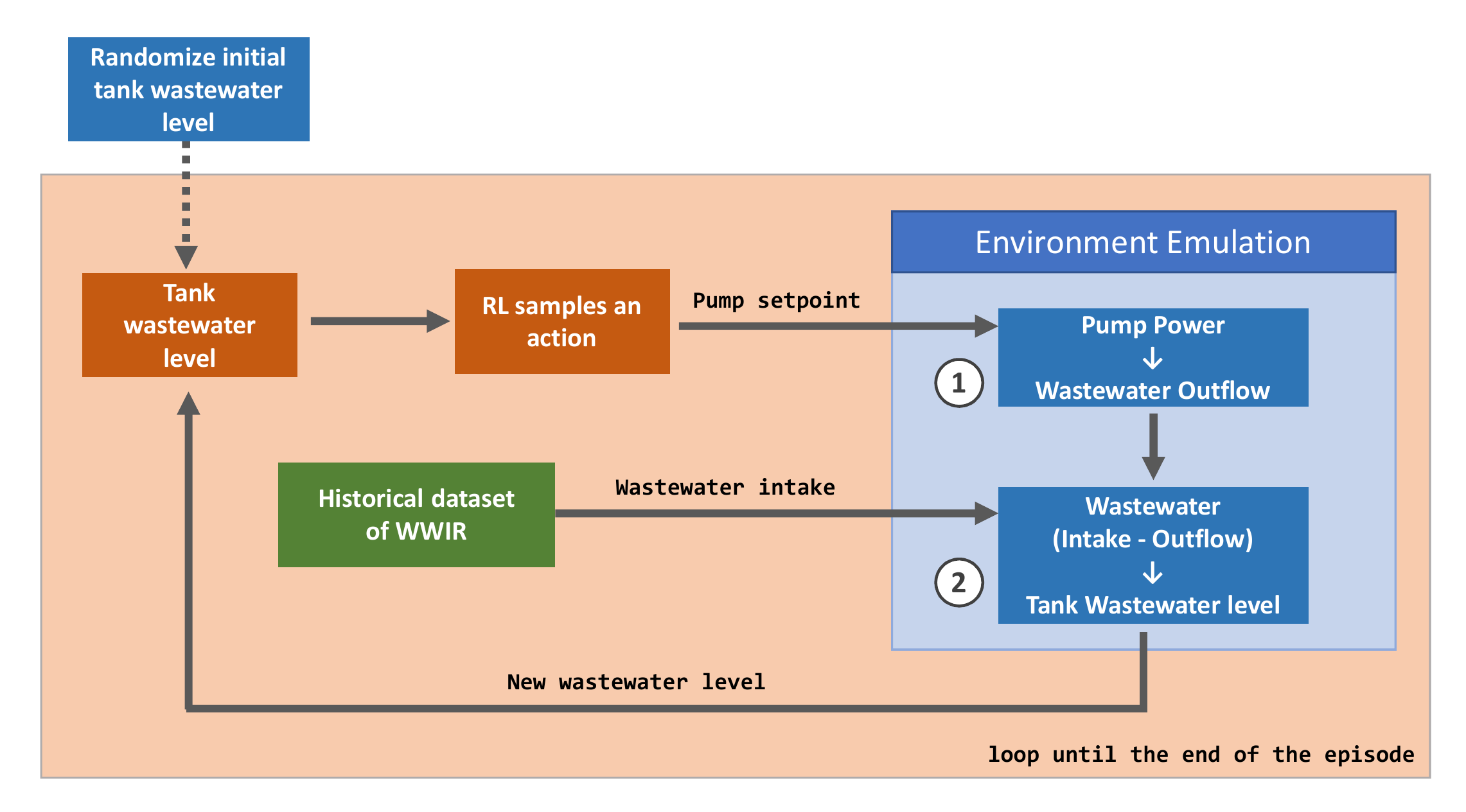}}
	\caption{Episodic training of the control policy: {\color{black} interaction, for each episode, between the control policy (reinforcement learning) and the emulated environment with supervised learning.}}
	\label{fig:emulation_framework}
\end{figure}

Figure~\ref{fig:emulation_framework} depicts the interaction, for each episode, between the control policy and {\color{black}the} emulated environment.

{\color{black} At the start of the episode, the initial level of the wastewater tank is randomly initialized, while for the remaining steps of the episode the wastewater level is a direct result of the pump operation, i.e. higher pumping power results in a lower level, while reducing the pumping operation leads to an increase in the tank level. The pumping operation (i.e., power set-points) are defined by the RL control policy, represented by a neural network that uses the state vector as the network input. The selected action is applied to the emulated environment: the first model translates pumping power into wastewater outflow rate, while the second one combines the wastewater outflow rate with the WWIR to emulate the change in the tank level. This updated value is then used at the start of the next step and the cycle is repeated.}

The following sections present {\color{black}the} details of the two models used to emulate the physical system, which are based in statistical learning algorithms.

\subsubsection{Modeling Wastewater Outflow Rate}
\label{sec:wwout}

Knowing the operational set-point of each individual pump (selected by the control policy) and the current level of wastewater of the tank, it is possible to estimate the amount of wastewater that will be pumped from the tank. The statistical learning algorithm GBT was used to model this relationship between pump power and wastewater outflow. For this model{\color{black},} the following input features were used: 

\begin{itemize}
\item Individual pump power set-point [float - kW] \\[-30pt]
\item Active pumps [binary] \\[-30pt]
\item Number of pumps active [integer] \\[-30pt]
\item Total power consumption [float - kW] \\[-30pt]
\item Tank water level [float - meters]
\end{itemize}

The data-driven approach allow{\color{black}s} us to model the pumping station environment{\color{black},} simply by using historical data. However, this particular dataset contained some inconsistencies due to noisy readings {\color{black}and} also due to a safety mechanism implemented in the pumping station. When the water level is above 7.2 meters{\color{black},} a portion of the wastewater is routed to the output of the wastewater, skipping some treatment stages. However, the SCADA system continues to collect data during these instants, leading to incorrect records. The first issue was minimized by applying a re-sampling technique which decreases the time resolution of the dataset from an average of 2 minutes for each reading to 4 minutes, reducing the number of samples but smoothing the records. For the second issue, all records registered when the wastewater level was above 7.2 meters were excluded from the learning dataset of this model.

\begin{figure}[H]
	\centerline{\includegraphics[scale=0.3]{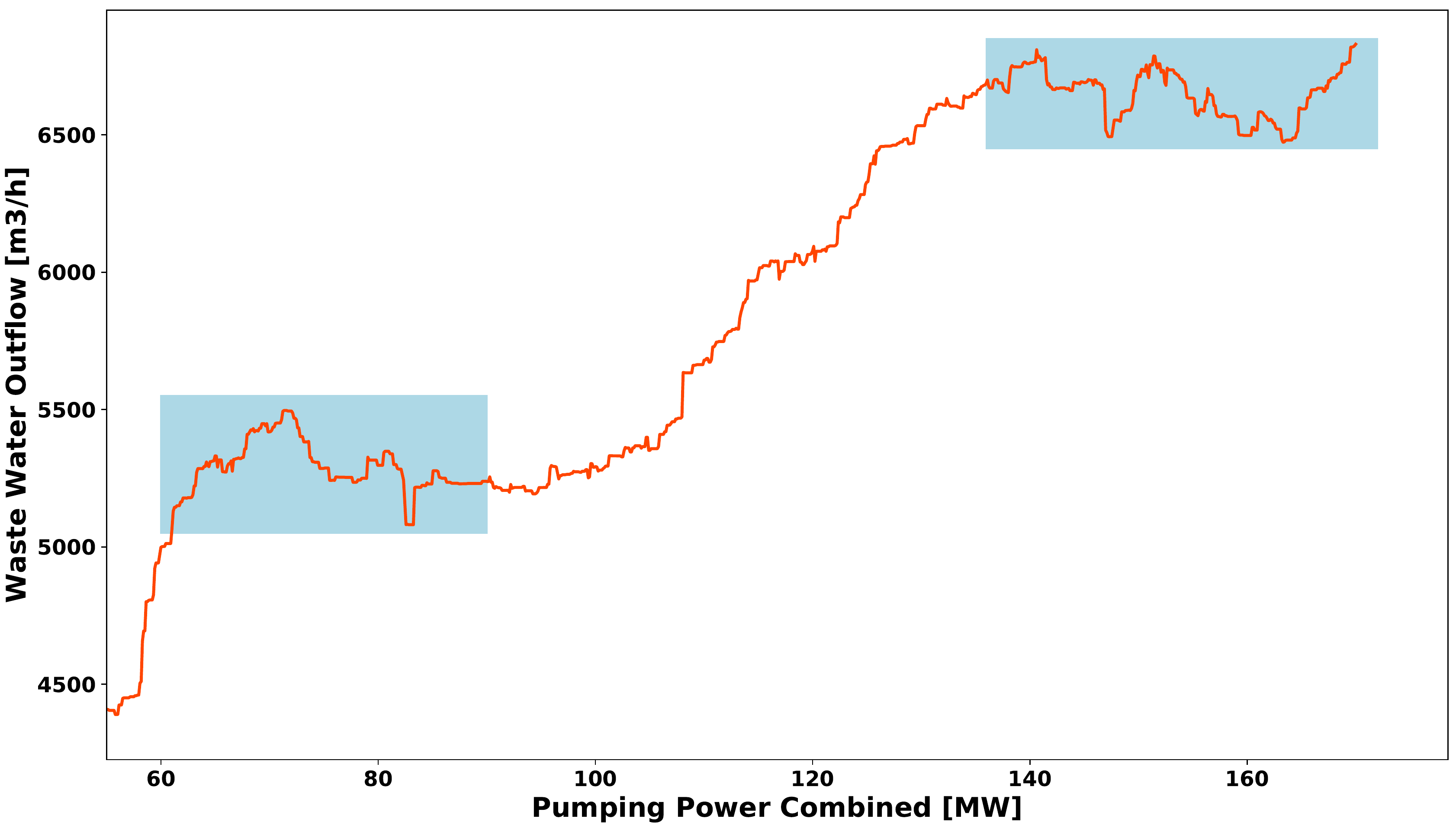}}
	\caption{{\color{black} Functional relation between wastewater outflow and pumping power. The reinforcement learning algorithm may present low generalization if it exploits only the shaded zones, leading to performance differences between environment simulation and operational deployment.}}
	\label{fig:EE_out_model_overfitted}
\end{figure}

Trying to optimize every hyper-parameter of the GBT model would allow us to improve the model's performance in the learning dataset, which typically is a good approach for a regression problem (note that we are not {\color{black}generating} forecasts). However, in a dataset with noise and incoherent readings, it would result in a model that does not accurately represent the actual physical system. Furthermore, as illustrated by Figure~\ref{fig:EE_out_model_overfitted}, if the RL algorithm discovers an operational range like the ones shaded, it will exploit this zone over and over{\color{black},} leading to a different performance when moving from the simulation environment to the operational deployment. 

The analysis of typical wastewater pumping curves allowed {\color{black}us} to conclude that the relationship between power, outflow and tank level is represented, typically, by a monotonic increasing curve, in contrast with the curve displayed in Figure~\ref{fig:EE_out_model_overfitted}, which shows neither of these characteristics. The integration of this domain knowledge into the data-driven approach has the potential to achieve a better modeling of the pumping station, and {\color{black}it} can be achieved by tweaking the hyper-parameters of the GBT training process. Decreasing the maximum depth of the individual regression estimators {\color{black}and} limiting the number of nodes in the gradient boost tree, forces the model to generalize better and avoids the excessive overfitting to the training dataset. Despite achieving a lower error metric when applied to the historical dataset, the model is able to better represent the physical system and, therefore, it is easier to be moved from simulation to physical application (requiring just a few interactions with the physical environment to calibrate the RL agent).

\begin{figure}[H]
	\centerline{\includegraphics[width=1\textwidth]{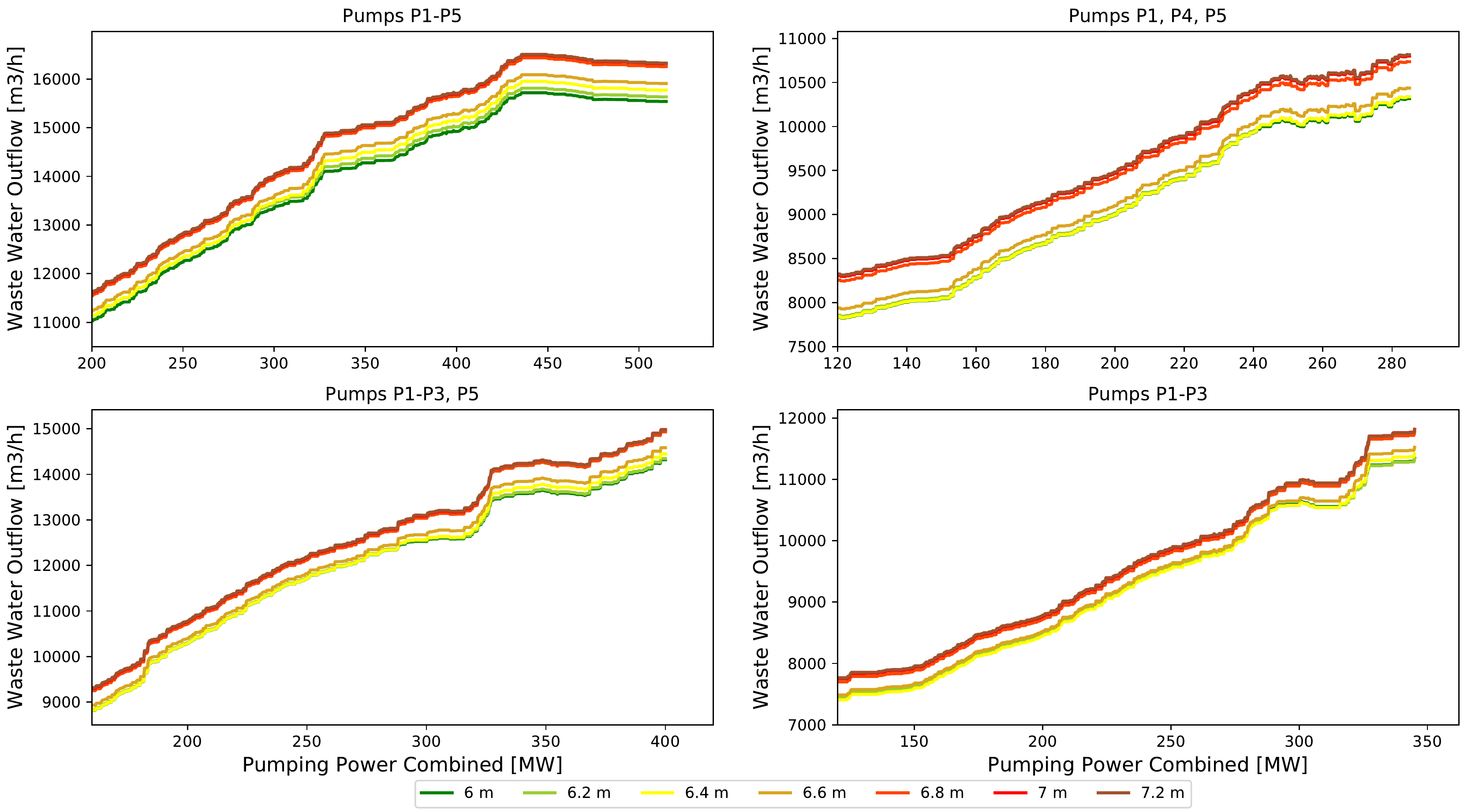}}
	\caption{{\color{black} Wastewater outflow as a function of the total power consumption and the current level of the tank, extracted from the fitted gradient boosting tree model in four different operating scenarios.}}
	\label{fig:EE_out_by_water_level}
\end{figure}

Figure~\ref{fig:EE_out_by_water_level} depicts, in four different operating scenarios, the results of the fitted model in which it is possible to observe the amount of wastewater outflow as a function of the total power consumption and the current level of the tank. One can verify the impact of the tank level in the efficiency of the pumps, i.e. {\color{black}the} higher the level, {\color{black}the} lower is the required power consumption. It is also possible to observe that the new GBT model outputs pumping curves with monotonic and increasing behavior.

The GBT model showed a MAE of 350 $m^3/h$ ({\color{black}2.43\% of NMAE\footnote{Normalized MAE: MAE / max WWIR value}}) and a root mean squared error (RMSE) of 478 $m^3/h$ ({\color{black}3.31\% of NRMSE\footnote{Normalized RMSE: RMSE / max WWIR value}}) in an out-of-sample dataset. 

\subsubsection{Modeling Reservoir Wastewater Level}
\label{sec:reser_level}

With {\color{black}the} information about the amount of WWIR entering the tank (by acquiring sensor data from the SCADA system) and knowing the amount of wastewater being pumped as {\color{black}a} result of the pumps operation (from the previous model - section \ref{sec:wwout}), it is possible to model the change in the tank level. 

Given the physical characteristics of the tank, this relationship can be modeled using a multi-linear regression model. The features used as input were: current wastewater tank level, WWIR, wastewater outflow rate and the difference between intake and outflow. Using the linear regression method from the Python \textsc{scikit-learn} library \cite{Pedregosa2011}{\color{black},} the model achieved a MAE of 4.32 centimeters ({\color{black}0.54\% of NMAE\footnote{Normalized NMAE: MAE / max tank level value}}) and a RMSE of 6.74 centimeters ({\color{black}0.84\% of NRMSE\footnote{Normalized RMSE: RMSE / max tank level value}}). 

\subsection{Control Policies}\label{sec:control_policies}

The control policy relies on RL to optimize the operation of the WWPS. A standard RL setup consists {\color{black}in} an agent interacting with an environment in discrete timesteps. At each timestep, the agent receives an observation $s_t$ (characterization of the state), takes an action $a_t$ and receives a scalar reward $r_t(s_t, a_t)$.

The agent behavior is defined by a policy $\pi_{\theta}(a|s_t)$ which samples an action given the current state and the policy parameters $\theta$. The aim of the agent is to maximize the sum of expected future rewards discounted by $\gamma$: $\mathbb{E}_{\pi}[R_t] = \mathbb{E} \big[\sum^{\infty}_{i\geq0} \gamma^i r(s_{t+i}, a_{t+i})\big]$.

\subsubsection{Agent}\label{sec:control_agent}

The RL algorithm used to train the control policy is the Proximal Policy Optimization (PPO)~\cite{PPO_2017}. PPO is a policy gradient method, which alternates between sampling data through interaction with the environment and optimizing a surrogate objective function using stochastic gradient ascent. 

The above-mentioned PPO algorithm typically uses an advantage estimator{\color{black},} based on the Generalized Advantage Estimation (GAE) \cite{GAE2015}, as described in Equation~\ref{eq:gae}, where $V(s_t)$ is the estimated value at the state $s$ and instant $t$, and $\gamma$ is the discount factor. To calculate this advantage{\color{black},} it is necessary to collect experiences for $T$ time steps and calculate the discounted advantage for all these time steps, so with $T=200$ the advantage at the first-time step will be composed by all the discounted rewards until the last-time step. This can lead to volatile results due to the credit assignment problem. In this work, we used a truncated GAE, so only a small part of all the experiences receive{\color{black}d} the discounted advantage, in order to better account the characteristics of the state vector. Since the WWIR forecasts used have a lead time of 20 steps ahead, it makes sense to limit the discounted advantage to the same interval, i.e. $T=20$.

\begin{equation}
\hat{A}_t = -V(s_t) + r_t + \gamma r_{t+1} + \cdots + \gamma^{T-t+1}r_{T-1}+\gamma^{T-1}V(s_T)
\label{eq:gae}
\end{equation}

The policy trained by PPO is represented by a neural network (NN) with two layers of 64 neurons and using the rectified linear unit as the activation function. The NN receives{\color{black},} as input{\color{black},} the state vector and outputs the parameters of a probability density function. Since the actions selected by the RL agent are within 0 and 1, corresponding to the set-point of each pump unit, a multivariate Beta distribution was used. Therefore, the NN output will be a pair of parameters from the Beta distribution ($\alpha$ and $\beta$) for each action. For this stochastic beta policy, only the cases where $\alpha, \beta > 1$ were considered, since solely in this domain the distribution is concave and unimodal.

During the training stage, actions are sampled from the probability density function (PDF)(Eq.~\ref{eq:beta_pdf}) in order to provide exploration, while during evaluation the stochasticity is removed and actions are selected as the mean of the PDF~(Eq.~\ref{eq:beta_mean}).

\begin{equation}
\label{eq:beta_pdf}
Beta_{PDF} = \frac{x^{\alpha - 1}(1-x)^{\beta - 1 }}{B(\alpha, \beta)}
\end{equation}
\begin{equation}
B(\alpha, \beta) = \frac{\Gamma(\alpha)\Gamma(\beta)}{\Gamma(\alpha+\beta)}
\end{equation}
\begin{equation}
\Gamma(z) = \int_{0}^{\infty} x^{z-1}e^{-x} \; dx
\end{equation}

\begin{equation}
\label{eq:beta_mean}
Beta_{mean} = \frac{\alpha}{\alpha + \beta}
\end{equation}

The complete configuration of the PPO agent is provided in Appendix B.

\subsubsection{Environment}\label{sec:control_environ}

The environment of the pumping station is described in section~\ref{sec:emulation}.

\subsubsection{State}\label{sec:control_state}

The state representation is a combination of the current snapshot of the environment and the WWIR forecasts presented in section~\ref{sec:forecasting}, namely:
\begin{itemize}
\item Current tank level \\[-30pt]
\item Pumps online (excluding units in maintenance or offline), which is a binary vector with the pump status  \\[-30pt]
\item Current WWIR \\[-30pt]
\item Current set-point for each individual pump \\[-30pt]
\item Probabilistic forecasts of WWIR: 25\%, 50\% and 75\% quantiles for 20 steps ahead  \\[-30pt]
\end{itemize}

\subsubsection{Actions}\label{sec:control_action}

The actions sampled from the control policy are the power set-points for each individual pump.

\subsubsection{Reward}\label{sec:control_reward}

The reward function quantifies the performance of the control policy, acting as the only feedback that allows the control policy to learn the correct action for a given state. 

Equation~\ref{eq:reward} defines the reward function used in this problem. As observed, the reward is divided into two terms: i) tank level and ii) power consumption. $c1$ and $c2$ are constants which give relative weight to each one of the terms, and are ordered according to the objective of the problem. Since it is more important to avoid {\color{black}the} overflow of the tank than to decrease the energy consumption: $c1 > c2$. The following sub-sections detail both parcels of the reward function.

\begin{equation}
r_t = c1 \cdot r_{WWlevel_t} + c2 \cdot r_{power_t}
\label{eq:reward}
\end{equation}
It is important to stress that the values $c1$ and $c2$ should be defined by the end-user according to its requirements. A sensitive analysis of these parameters is provided in section \ref{sec:reward_coefs}.

\begin{description}[leftmargin=0.6cm,font=\normalfont]

	\item [\textit{Wastewater Reservoir Level, $r_{WWlevel_t}$}] \hfill \\[5pt] 
	The first term rewards the policy for operating with the tank level within the admissible limits. If the level is between 3m and 7.2m, a reward of $R^{+}$ is given, otherwise a reward of $R^{-}$ is issued. An empirical analysis showed good results with $R^{+} = 3$ and $R^{-} = -600$. It is relevant to stress that it is important to have a much severe penalty than a reward, since the first objective is to operate within the tank limits. Mathematically{\color{black},} it is translated into:
	
\begin{equation}
  r_{WW_{level,t}}=\begin{cases}
               R^{+},\;\;\; h \in [3.0, 7.2]\\
               R^{-},\;\;\; h \in [0, 3.0[ \;\; \vee \;\; h \in ]7.2, 8.0]
            \end{cases}
\end{equation}
	
	\item [\textit{Power Consumption, $r_{power_t}$}] \hfill \\[5pt] 
	Since the control strategy aims to decrease the power consumption of the WWPS, for each timestep $t${\color{black},} a penalty proportional to the total installed capacity of the pumping station is applied as follows: 
    
    \begin{equation}
     r_{power_t} = \frac{\sum_{i}^{} P_{i,t}}{P_{inst}}
    \end{equation}
where $P_{i,t}$ is the power consumption of the i-th pump for timestep $t$ and $P_{inst}$ the total capacity of the WWPS.
	
\end{description}

\subsubsection{RL Agent Learning Process}\label{sec:control_train}

The training process consists {\color{black}i}n performing a set of actions for several iterations, using the simulated environment and two sets of episodes: one for training and the other to continuously assess the performance of the control policy.

The policy NN is initialized by assigning random values to its weights and bias. Then an iterative learning process takes place until the predefined number of iterations is reached. Each iteration is divided into two phases: training and test.

The training process randomly select{\color{black}s} an episode{\color{black}, which enables} the RL agent to interact with the simulated environment for $T$ timesteps. In this stage, both the policy and the environment are stochastic in order to provide exploration. Since the policy is represented by a beta distribution, the actions are obtained taking samples from the PDF. In the environment, the stochasticity is provided by randomizing the wastewater level in the tank at the start of each episode. In each interaction (step){\color{black},} the control policy receives as input the state of system (as described in Section~\ref{sec:control_state}) and chooses an action representing the operational set-point for each one of the pumps. The action is applied to the environment which leads the system to a new state and {\color{black}to} the emission of a reward. For each step, the state transition vector (observation, action, reward) is collected, and this episodic roll-out ends at the terminal stage of an episode (either by reaching the maximum number of steps or by getting an alarm). Afterwards, the collected vector is used to train the RL agent using the PPO algorithm, altering the policy parameters (weights and bias of the neural network). 

The recently updated policy is then used in the second stage and is used to assess its performance in a small subset of the available testing episodes. In this stage{\color{black},} the stochasticity is removed to allow reproducibility between iterations, thus leading to a fairer comparison and analysis. The statistics collected during this stage (average reward, number of alarms and energy consumption) are used to evaluate the learning process. This concludes an iteration of the learning process and the policy restarts the collection of the transitions vector.

\section{Numerical Results}\label{sec:results}

The energy optimization strategy was applied to the WWPS described in section~\ref{sec:process_description}.

Using the historical records collected by the station's SCADA system, a set of episodes were created for evaluating the control performance, i.e. 80\% were chosen to train the RL agent{\color{black},} while the remaining 20\% were used as the test set. Due to some missing values in the original dataset, the episodes have variable length{\color{black}s} in order to maximize the amount of continuous steps.

A minimum of 200 timesteps was enforced per episode, while the maximum was left uncapped, leading to some episodes with more than 3000 timesteps. Each episode contains information about the number of online pumping units (excluding the ones in maintenance or unavailable), current and forecasted WWIR. This information is discretized in 2-minute steps.

The numerical results presented below were obtained by applying the RL {\color{black}agent to all episodes in the testing set (i.e., 65 episodes), encompassing more than 64,000 steps for a total of 90 days of data.}

Four scenarios were considered to measure the performance of the data-driven optimization strategy: 
\begin{itemize}
\item Current operating rules 
\item Strategy without WWIR forecasts
\item Strategy with WWIR forecasts
\item Strategy with perfect WWIR forecasts (i.e. observed values)
\end{itemize}

The first scenario is used as the benchmark model{\color{black},} since it corresponds to the historical records of the current operation of the WWPS. The other three scenarios aim to assess the improvement of the RL agent{\color{black},} in comparison with current control operation and also to compare the predictive aspect of RL agent by evaluating an agent with access to WWIR forecasts. The perfect forecast scenario establishes an upper bound for the electrical energy saving gains.

The following sections show a set of numerical analysis considering these four scenarios. It is important to underline that the RL learning process has an inherent instability. As mentioned before, for each learning iteration{\color{black},} the control policy is applied to the test set and the relevant performance indicators are collected. Furthermore, instead of considering the best iteration of all the learning process, it is considered the last 5000 iterations to smooth the variability. The results are presented as a set of quantiles (25\%, 50\% and 75\%) to better illustrate the variability in performance from {\color{black}one} iteration to {\color{black}the other}.

\subsection{Improvement in the Number of Alarms}

\begin{figure}[H]
	\centerline{\includegraphics[width=1\textwidth]{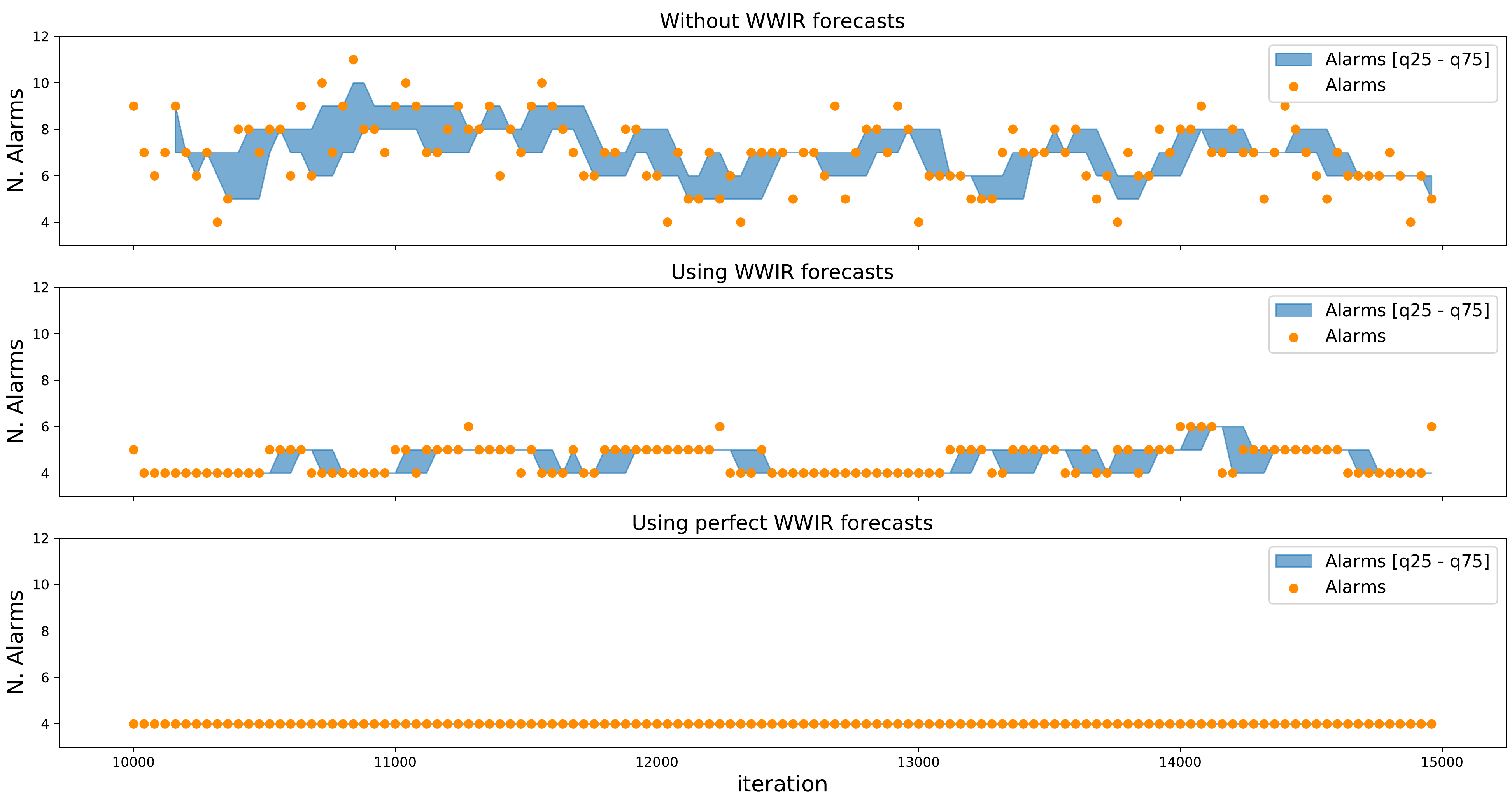}}
	\caption{Number of alarms (wastewater level above 7.2m) {\color{black} using the reinforcement learning method with and without wastewater intake rate forecasts. The depicted dots represent the number of alarms while the shaded area indicates the 25\% and 75\% quantiles of the same record.}}
	\label{fig:alarms_quantiles}
\end{figure}

The WWPS has preventive measures to avoid spillage of wastewater during periods of extreme WWIR. This section studies the number of alarms triggered in the four scenarios under evaluation. Results are depicted in Figure~\ref{fig:alarms_quantiles} and Table~\ref{tab:n_alarms}. 

Figure~\ref{fig:alarms_quantiles} shows the number of alarms registered when operating the WWPS using the RL control for the past 5000 iterations of the training process. The top plot shows the control results without WWIR forecasts in the state vector, the middle plot shows the predictive control with probabilistic WWIR forecasts, and the bottom one indicates the alarm performance considering perfect information of WWIR. Without WWIR forecasts{\color{black},} the number of alarms ranges from 4 to 11, while{\color{black},} with the predictive control{\color{black},} the number of alarms is kept between 4 and 6 and, as {\color{black}depicted} in the plot, with much less variability. With perfect forecast for the WWIR, the predictive control maintains an alarm number of 4. As {\color{black}it} will be discussed later, 4 alarms are the minimum possible number, since these alarms occurred at the start of the episode, therefore unavoidable for the RL control strategy.

Table~\ref{tab:n_alarms} shows a comparison between the RL control strategy and the current operational procedure. Considering the episodes under evaluation, a total of 1671 occurrences were originally registered {\color{black}for a} wastewater level above the first alarm trigger (7.2 meters). Despite the previous numbers, the second alarm (at 8 meters) was never triggered, registering a maximum level of 7.98 meters.

The RL approach was able to significantly reduce the number of alarms triggered, for both the predictive and non-predictive strategies. Without WWIR forecasts, the RL strategy registered an average of 7 alarms and, by including the probabilistic forecasts{\color{black},} this number decreased to 4.55 alarms in average, for the last 5000 training iterations.

\begin{table}[H]
	\centering
	\caption{Number of reservoir level alarms triggered {\color{black}by the RL control strategy and the current operational procedure. Two critical alarm levels are considered: a first one at 7.2 meters (initiates a preventive action) and the other at 8 meters when overflow occurs.}}
	\begin{tabular}{l | c c c | c  }
		\toprule
		\toprule
		 &  \multicolumn{3}{c|}{Alarms 7.2m} & Alarms 8m  \\
		 & q25\% & q50\% & q75\% & n. occur \\
		\hline
		Current operating rules    & \multicolumn{3}{c|}{1671} & 0 \\
		RL without WWIR forecasts & 6.47  & 7.05 & 7.54 & 0  \\
		RL with WWIR forecasts    & 4.36  & 4.55 & 4.68 & 0 \\
        RL with perfect WWIR forecasts    & 4.0  & 4.0 & 4.0 & 0 \\
		\bottomrule
		\bottomrule
	\end{tabular}%
	\label{tab:n_alarms}%
\end{table}

\begin{figure}[H]
	\centerline{\includegraphics[scale=0.45]{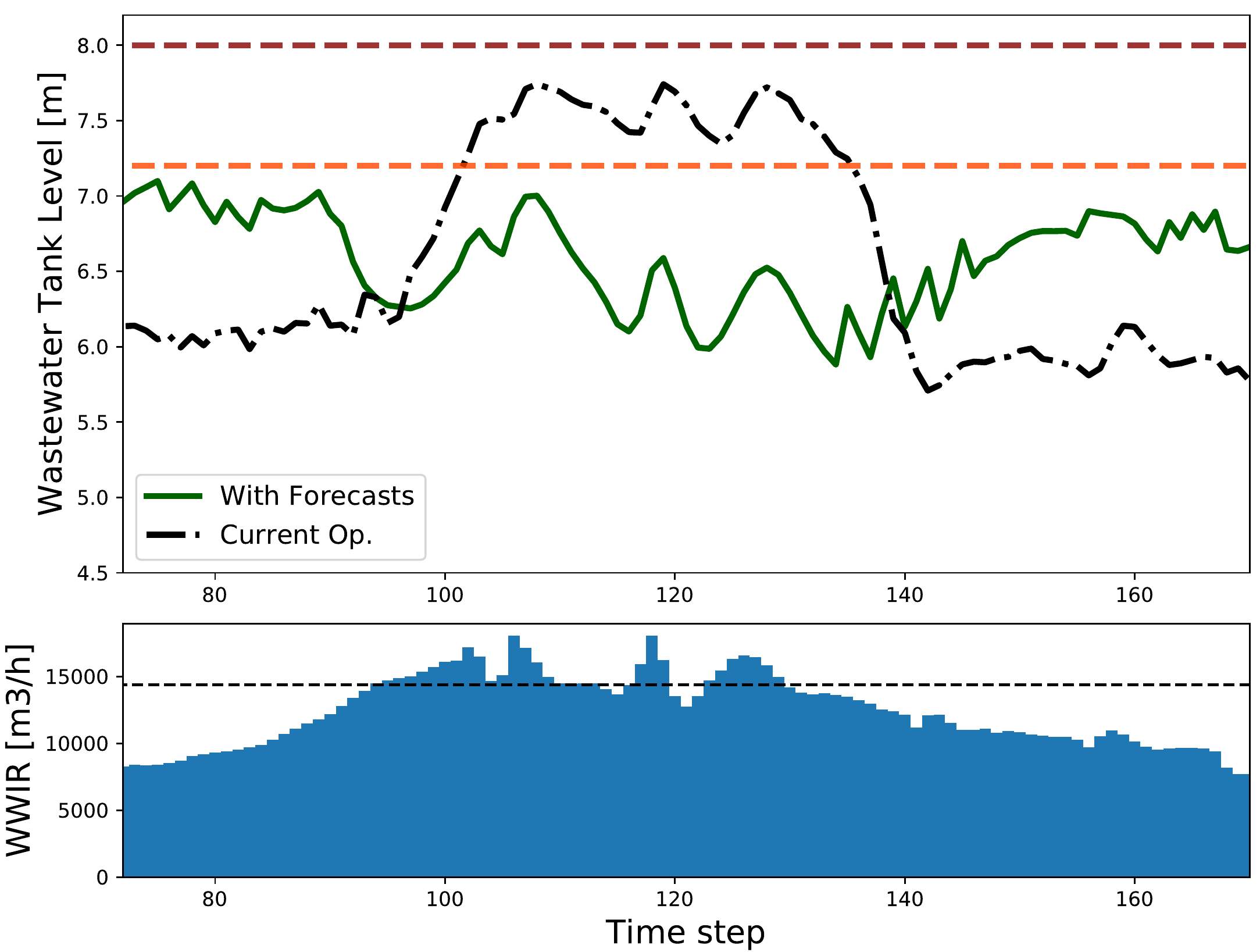}}
	\caption{{\color{black}Illustrative example showing that the predictive control is capable of anticipating an increase in the wastewater intake rate and of adjusting the pumping operation accordingly. The top plot depicts the tank level for predictive control and current operating rules. The bottom plot shows the observed wastewater intake rate.}}
	\label{fig:prevent_alarms}
\end{figure}

Providing a set of probabilistic forecasts to the RL agent adds the possibility to anticipate changes in the WWIR and adjust the pumping operation accordingly. Figure~\ref{fig:prevent_alarms} illustrates this capability. The control strategy operates at a tank level close to the first alarm trigger in order to optimize the electrical energy consumption, but when the forecasts show an expected increasing rate of WWIR{\color{black},} the pumping activity is increased to provide a sufficient buffer to accommodate the expected increase of the tank level. {\color{black}The control strategy with the current operating rules} is not quick enough to adapt to the changing rates and the first alarm is triggered during {\color{black}a} time period. 

\begin{figure}[H]
	\centerline{\includegraphics[scale=0.35]{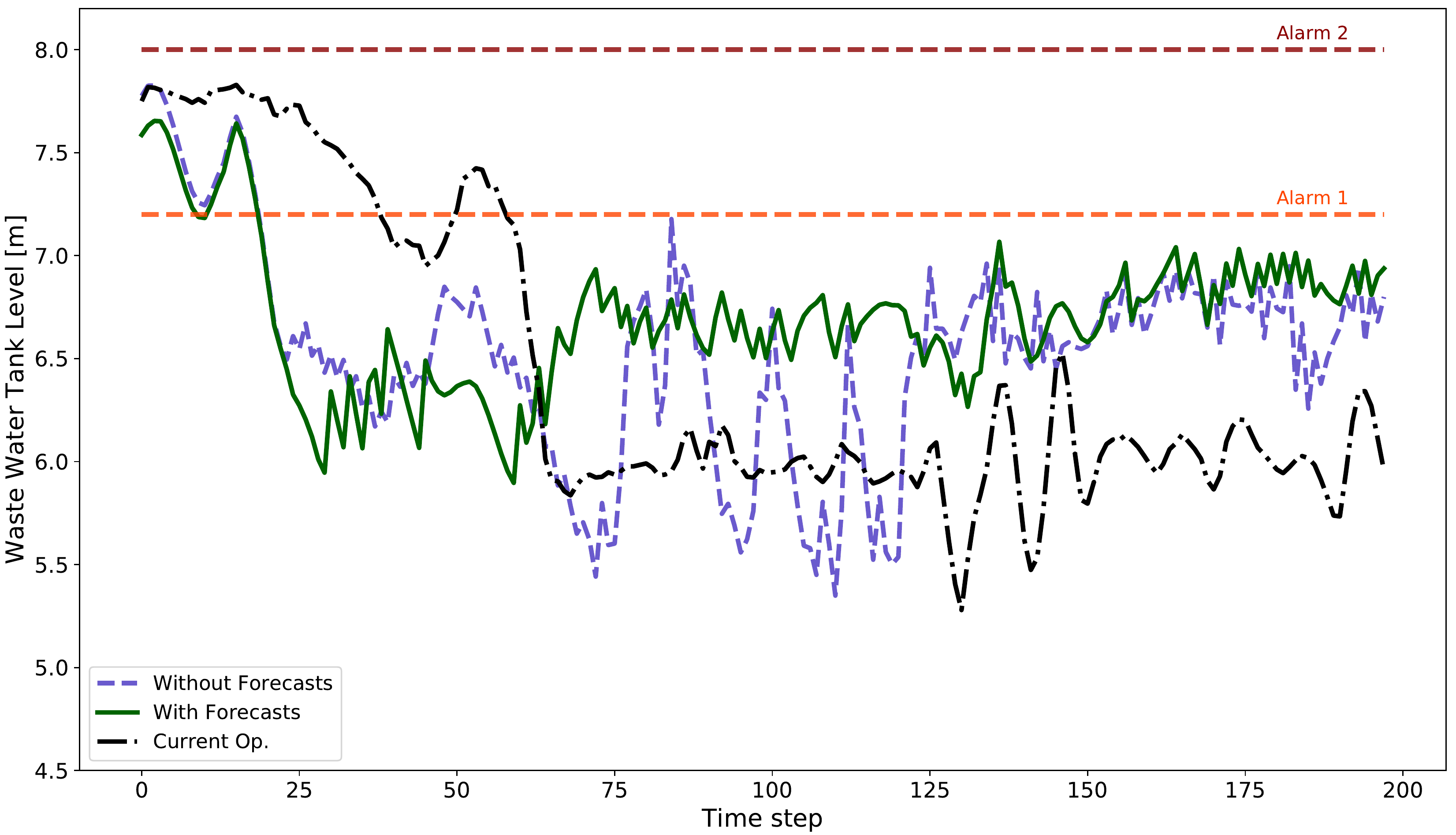}}
	\caption{{\color{black}Operating situation where the initial tank level was already above the maximum level trigger and it was not possible to avoid the alarm.}}
	\label{fig:alarms_starting}
\end{figure}

Despite the significant improvement in performance of the RL control strategy, in comparison with the benchmark operation, it still registers a few alarms. However, some of these alarms occurred in unavoidable situations. After dividing the historical dataset into segments, some of the episodes start{\color{black}ed} with the tank level already above the first alarm. In fact, four of the registered alarms occurred in the first and second steps of an episode, which are unavoidable for the control strategy (even with perfect WWIR forecasts). Figure~\ref{fig:alarms_starting} depicts one of those situations{\color{black},} where the initial tank level was already above the alarm trigger and{\color{black},} therefore{\color{black},} unavoidable. However, the RL {\color{black}control strategy} was able to significantly reduce the amount of timesteps with tank above 7.2 meters. 

\subsection{Improvement in Electrical Energy Consumption}

\begin{figure}[H]
	\centerline{\includegraphics[width=1\textwidth]{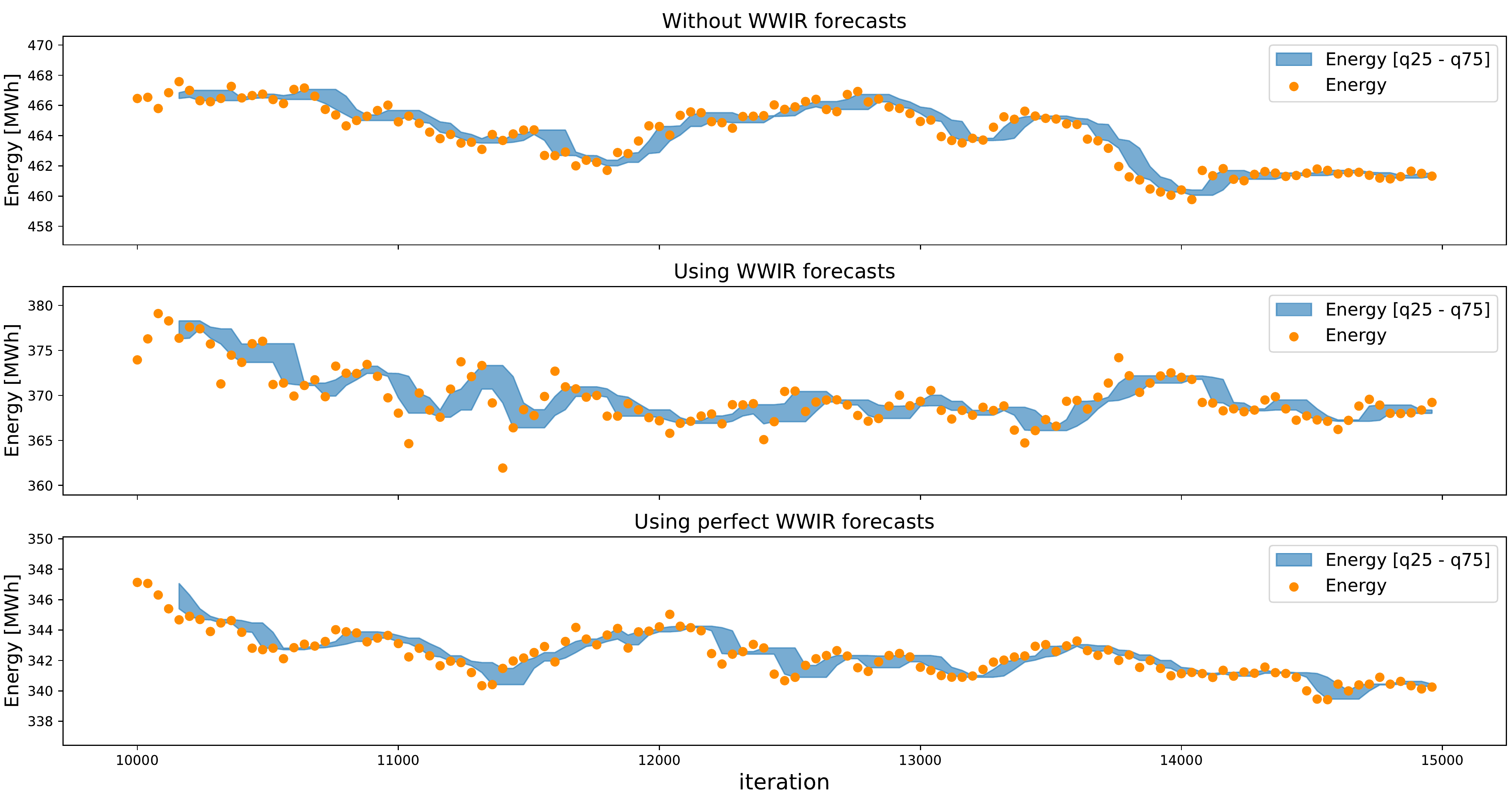}}
	\caption{{\color{black}Electrical energy consumption obtained with the reinforcement learning control for three levels of information: 1) without wastewater intake rate forecasts; 2) with probabilistic forecasts; 3) using perfect forecasts.}}
	\label{fig:energy_quantiles}
\end{figure}

\begin{figure}[H]
	\centerline{\includegraphics[scale=0.45]{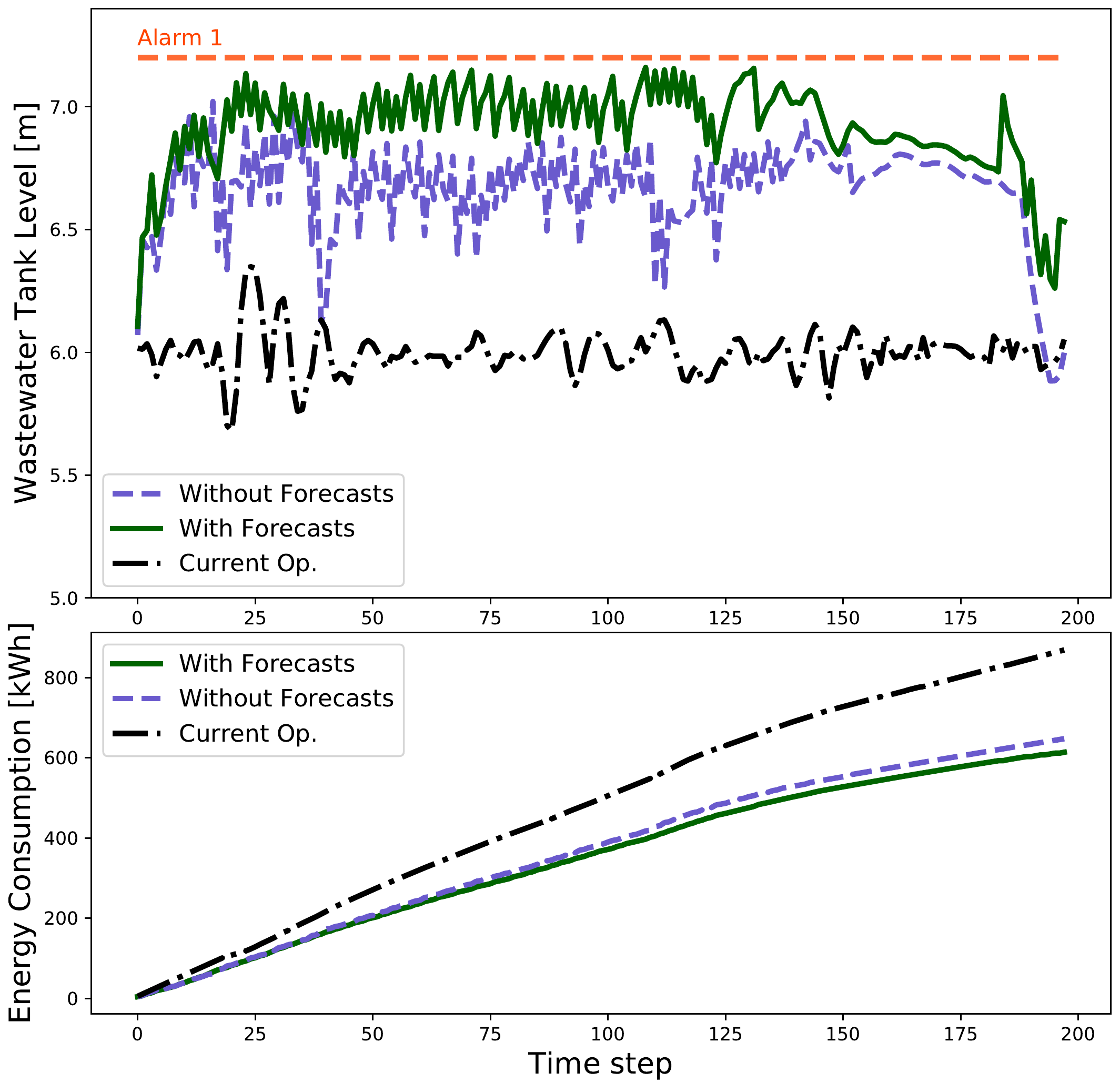}}
	\caption{{\color{black}Wastewater level and energy consumption for one episode under three different test scenarios: current operating rules, with and without wastewater intake rate forecasts. The top plot shows the tank level while the bottom plot depicts the cumulative electrical energy consumption.}}
	\label{fig:results_energy_opt}
\end{figure}

This section studies the ability of the proposed control strategy to reduce the electrical energy consumption of the WWPS process. Following the methodology presented in the previous section, Figure~\ref{fig:energy_quantiles} depicts the comparison between the predictive and non-predictive RL control, while Table~\ref{fig:results_energy_opt} shows the results for the 
absolute values of electrical energy consumption for the four scenarios under evaluation{\color{black},} plus the improvement of the RL control in comparison with the current operating rules. 

The results in Figure~\ref{fig:energy_quantiles} show that the non-predictive control registered a cumulative energy consumption ranging between 459~MWh and 469~MWh, while the predictive control was able to operate with significantly less electrical energy needs, obtaining values between 362~MWh and 379~MWh. The scenario with WWIR perfect forecasts showed an electrical energy consumption between 340~MWh and 348~MWh.

From Table~\ref{tab:energy_consumption}{\color{black},} it is possible to observe that the current operating rules required a cumulative electrical energy consumption of 379.8~MWh. In comparison, the non-predictive strategy required more 22\% than the current operation, while using WWIR probabilistic forecasts as input resulted in an added value{\color{black},} since the electrical energy needs {\color{black}were} reduced by almost 3\%. It is important to underline that with perfect WWIR forecasts, the control strategy was able to achieve an improvement of 9.95\%.

\begin{table}[H]
	\centering
	\caption{{\color{black}Total} electrical energy consumption in the three scenarios: current operating rules, with and without WWIR forecasts.}
	\begin{tabular}{l | c c c | c }
		\toprule
		\toprule
		 &  \multicolumn{3}{c|}{Energy Consumption [MWh]} & Improvement \\
		 & q25\% & q50\% & q75\% &  q50\% \\
		\hline
		Current operating rules    & \multicolumn{3}{c|}{ 379.8 } & --  \\
		RL without WWIR forecasts & 463.7  & 464.0 & 464.3 & -22.19\%  \\
		RL with WWIR forecasts    & 369.0  & 369.7 & 370.5 & 2.66\%   \\
        RL with perfect WWIR forecasts    & 342.0  & 342.32 & 342.57 & 9.95\%   \\
		\bottomrule
		\bottomrule
	\end{tabular}%
	\label{tab:energy_consumption}%
\end{table}

The amount of energy consumption required to operate the WWPS has a direct relation with the average wastewater tank level, i.e. a higher level requires less power to pump the same amount of wastewater. Table~\ref{tab:tank_level} shows the average and standard deviation of wastewater tank level for the four scenarios under evaluation. As expected, with the WWIR forecast for 20 steps ahead, the control strategy is able to operate the facility at almost 30 cm higher than the current operating rules, which explains the electrical energy savings. Furthermore, the scenario with perfect WWIR forecasts obtained an average tank level of 6.58~m.

\begin{table}[H]
	\centering
	\caption{Wastewater tank level {\color{black}(mean and standard deviation)} in the three scenarios: current operating rules, with and without WWIR forecasts.}
	\begin{tabular}{l | c c}
		\toprule
		\toprule
		 &  \multicolumn{2}{c}{Wastewater level}\\
		 & mean [m] & std [m]\\
		\hline
		Current operating rules    & 6.05 & 0.39\\
		RL without WWIR forecasts & 6.06 & 0.35 \\
		RL with WWIR forecasts    & 6.36 & 0.21 \\
        RL with perfect WWIR forecasts    & 6.58 & 0.15 \\
		\bottomrule
		\bottomrule
	\end{tabular}%
	\label{tab:tank_level}%
\end{table}

Figure~\ref{fig:results_energy_opt} depicts an episode where it is possible to observe the lower electrical energy {\color{black}consumption} of the RL strategy in comparison to the current operating rules. In the current operation scenario, the pumps are operated in order to keep the tank level around 6 meters, while the RL strategies are free to choose the optimal operation point. As a result, the strategy which uses WWIR forecasts is able to operate very close to the first alarm trigger in order to reduce the electrical energy use. The {\color{black}inclusion of WWIR forecasts} provides the RL strategy with a snapshot of the following instants and allows an {\color{black}operation close to the alarm threshold} while maintaining the risk under control. 

\subsection{Trade-off Between Alarms and Electrical Energy Consumption}\label{sec:reward_coefs}

As discussed in Section~\ref{sec:control_reward}, in the reward function the coefficients $c1$ and $c2$ represent the weight assigned to both objectives of the control strategy: alarms and electrical energy consumption reduction, respectively. In this section, the impact of changing these values is analyzed.

Two scenarios were considered:

\begin{itemize}
\item \textbf{Scenario\textsuperscript{alarms}:} prioritizes the reduction on the number of alarms ($c1 = 1$; $c2 = 0.5$) and
\item \textbf{Scenario\textsuperscript{energy}:} emphasizes the electrical energy consumption reduction ($c1 = 0.5$; $c2 = 1$)
\end{itemize}

\begin{figure}[H]
	\centerline{\includegraphics[width=1\textwidth]{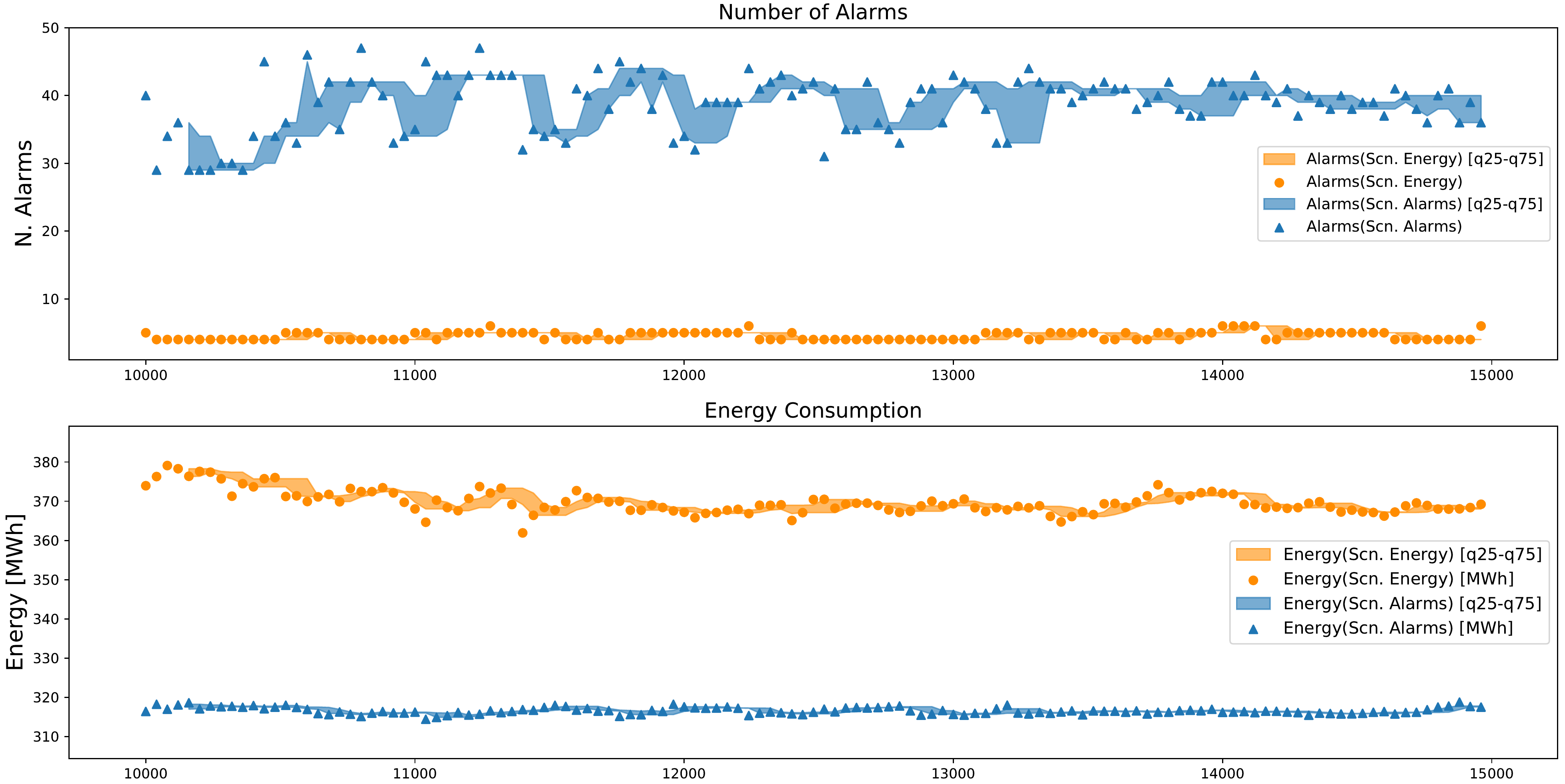}}
	\caption{Number of alarms versus electrical energy consumption in two different scenarios:{\color{black} 1) \textbf{Scenario\textsuperscript{alarms}} that prioritizes the reduction on the number of alarms; 2) \textbf{Scenario\textsuperscript{energy}} that emphasizes the electrical energy consumption reduction. The top plot depicts the number of alarms and the energy consumption is presented in the bottom plot.}}
	\label{fig:trade_off}
\end{figure}

Figure~\ref{fig:trade_off} presents a comparison between the two scenarios. In Scenario\textsuperscript{alarms}, the number of alarms is low, ranging between 4 and 7, while the other scenario reaches numbers between 29 and 48. By analyzing the electrical energy consumption, it is possible to detect a considerable difference between both scenarios, i.e. Scenario\textsuperscript{alarms} has a{\color{black}n} energy consumption ranging from 366~MWh to 377~MWh, while {\color{black}in} Scenario\textsuperscript{energy} {\color{black}ranges} between 315~MWh and 318~MWh.

Table~\ref{tab:trade_off} details the average values for alarms and electrical energy consumption for both scenarios and the current operating control rules, plus the improvement obtained with the RL control. As seen in Figure~\ref{fig:trade_off}, Scenario\textsuperscript{energy} obtains a lower energy consumption at the cost of more alarms. {\color{black}Putting this increase into} perspective, even with the record of more than 38 alarms, the Scenario\textsuperscript{energy} lead{\color{black}s} to a decrease of almost 98\% in the number of alarms when compared with the current operation. On the other hand, energy consumption decreases almost 17\% when compared with the current operating rules, while with the Scenario\textsuperscript{alarms} the decrease was only 2.66\%. Therefore, these results show that{\color{black},} with the proposed methodology, significant reduction in electrical energy consumption can be obtained with a minor increase in the number of alarms. In all cases, the improvement in both criteria was obtained compared to the current operating rules.

\begin{table}[H]
	\centering
	\caption{Comparison between the current operating rules and the two scenarios under evaluation: Scenario\textsuperscript{alarms} and Scenario\textsuperscript{energy}.}
	\begin{tabular}{l | c c | c c }
		\toprule
		\toprule
		 &\multirow{2}{*}{Energy [MWh]}&\multirow{2}{*}{Alarms}&\multicolumn{2}{c}{Improvement} \\
		 &  & & Energy & Alarms \\
		\hline
		Scenario\textsuperscript{alarms} & 369.7 & 4.55 & 2.66\% & 99.73\%\\
		Scenario\textsuperscript{energy} & 316.5 & 38.7 & 16.67\% & 97.68\% \\
		Current operating rules & 379.8 & 1671 & -- & --\\
		\bottomrule
		\bottomrule
	\end{tabular}%
	\label{tab:trade_off}%
\end{table}

\section{Conclusions}\label{sec:conclusions}

The present work proposes a data-driven optimization framework for the operation of a wastewater pumping station. This work explores machine learning techniques that produce probabilistic forecasts for the wastewater intake rate and construct a data-driven proxy for the physical system. By feeding a reinforcement learning algorithm with these forecasts and proxy models, a predictive control algorithm is implemented to optimize (i.e. minimize electrical energy consumption and alarms for tank level above limit) the operation of wastewater pumps by actuating in its active power set-points.

{\color{black}The proposed method showed the following advantages: i) {\color{black}it} mitigates several shortcomings that occur when using physics-based models{\color{black},} such as the need to have a good knowledge of the process (e.g., pump operating curves, parameters), the time required to build a physical model, and the limited availability of sensors - these features facilitate replication in similar systems and in other wastewater installations; ii) it enables continuous learning and self-adaption of pumps (e.g., different levels of wear and tear) due to learning capabilities that provided reinforcement learning when interacting with the real environment (system); iii) the predictive approach automatically anticipates the periods of high wastewater intake rate and avoids alarms (tank level above 7.2 meters).} 

Operating the wastewater pumping station optimally depends on the trade-off between the number of alarms and the electrical energy consumption of the facility. The results in a real-world wastewater pumping station for 90 non-consecutive days showed a higher performance both in terms of energy efficiency and alarms. If priority is given to the operation within the wastewater tank limits, the proposed control strategy {\color{black}is} able to nearly mitigate them, {\color{black}as only} 4 alarms {\color{black}were obtained, which} occurred in unavoidable situations (99\% less than the current operating rules). Furthermore, a decrease of almost 3\% of the overall electrical energy needs of the pumping station was also verified. On the other hand, {\color{black}when} more weight {\color{black}was} given to the electrical energy component, the control strategy registered 39 alarms (still meaning a 97\% decrease in comparison with the current procedures) while improving the energy efficiency in 17\%.

Finally, {\color{black}it is} important to mention that around 20\% of the operational costs in the company's {\color{black}wastewater treatment plants} were associated to the wastewater pumping stations. Therefore, the benefits of the proposed method could be translated into a significant reduction of the global energy consumption.

\section*{Acknowledgements}

The research leading to this work is being carried out as a part of the InteGrid project (\textit{Demonstration of INTElligent grid technologies for renewables INTEgration and INTEractive consumer participation enabling INTEroperable market solutions and INTErconnected stakeholders}), which received funding from the European Union’s Horizon 2020 Framework Programme for Research and Innovation under grant agreement No. 731218. 

The sole responsibility for the content lies with the authors. It does not necessarily reflect the opinion of the Innovation and Networks Executive Agency (INEA) or the European Commission (EC). INEA or the EC are not responsible for any use that may be
made of the information it contains.

\appendix

{\color{black}\section{Forecasting Results for Wastewater Intake Rate}}
\label{app:forecast_WWIR}
\setcounter{secnumdepth}{0}
\subsection{Forecasting Skill Metrics}

The forecasting skill of the wastewater inflow was evaluated for both point and probabilistic forecast. The quality of the probabilistic forecast was access{\color{black}ed} with the following metrics: calibration, sharpness and continuous rank probability score (CRPS). A completed description of these metrics is given in \cite{metricsPinson}, and the following paragraphs present a general description of these metrics.

Calibration measures the deviation between empirical probabilities (or long-run quantile proportions) and nominal probabilities{\color{black},} and is usually calculated for each quantile nominal proportion $(\tau)$.  Sharpness quantifies the degree of uncertainty in probabilistic forecasts, which numerically corresponds to comput{\color{black}ing} the average interval size between two symmetric quantiles (e.g., $10\%$ and $90\%$ with coverage rate $\gamma$ equal to 20 \%).

CRPS is {\color{black}a} unique skill score that provides the entire amount of information about a given {\color{black}method} performance and encompasses information about calibration and sharpness of probabilistic forecasts. The CRPS metric was adapted to evaluate quantile forecasts, as described in \cite{Taieb2016}).

Point forecasts, i.e. 50\% quantile in this work, were evaluated with the classical Mean Absolute Error (MAE). 

\subsection{Benchmark Models}

Typically used in the literature, the persistence model assumes that the forecast for $t+1$ is equal to the value observed in the previous time step $t$. Due to the strong correlation between time instants $t$ and $t-1$, a naive algorithm such as this one can be very hard to beat in terms of forecasting performance, especially for very {\color{black}short-term} time horizons and in dry periods with low variability.

In addition to persistence, LQR model conditioned to the period of the day (i.e., hour) is also considered as a second benchmarking model. With this benchmark method, we allow the distribution to change for each period of {\color{black}the} day. This method will be denoted $CondbyHour$.

\subsection{Variables Importance}\label{app:variables_imp}

\begin{figure}[h]
	\centerline{\includegraphics[scale=0.7]{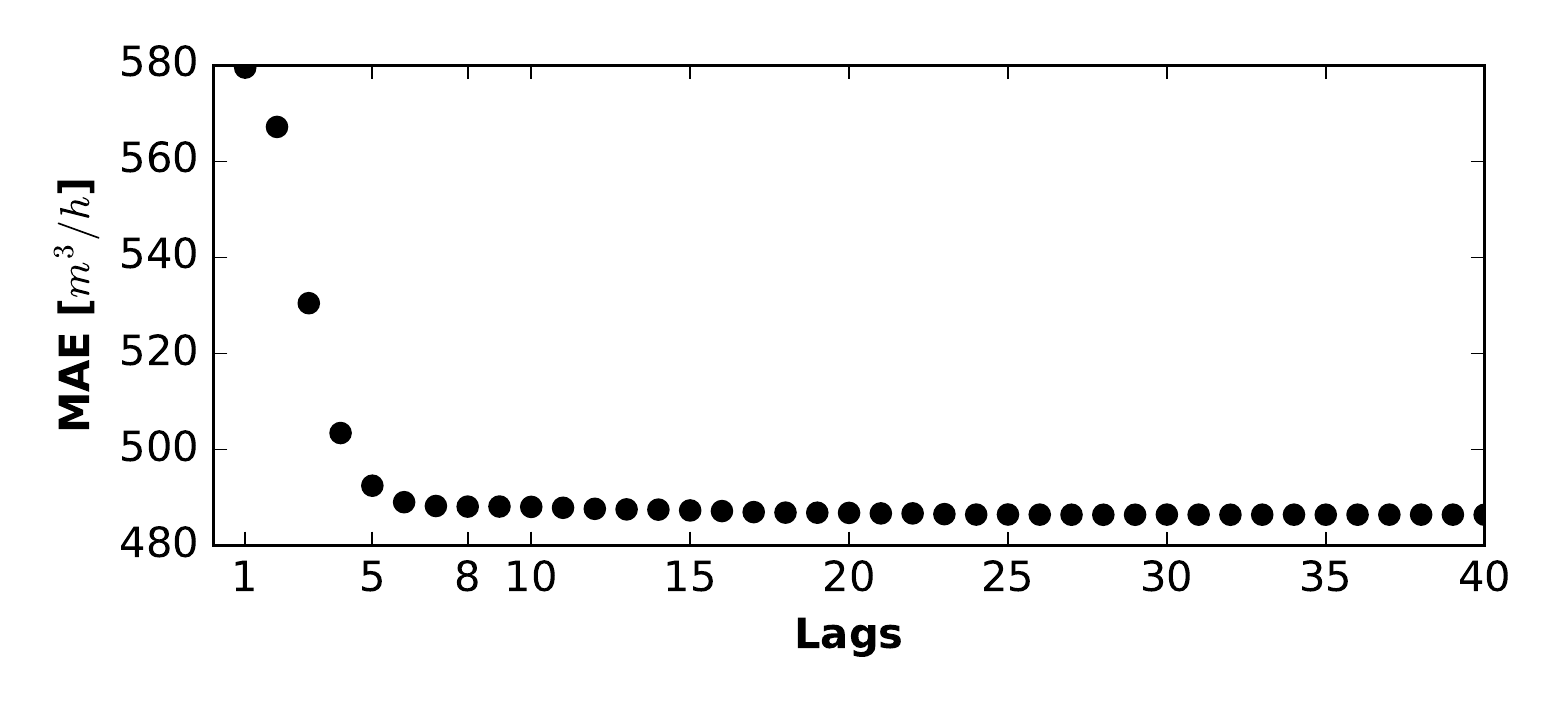}}
	\caption{Mean absolute error (MAE) as a function of the number of lags.}
	\label{fig:lag}
\end{figure}

This subsection evaluates whether or not the input features improve the forecasting skill, measured with the MAE and CRPS. This evaluation is performed only for the first lead time ($t+1$) using the LQR. 
The choice o{\color{black}f} the number lags was made by calculating the MAE of different models having {\color{black}a} different {\color{black}number} of lags as explanatory variables (parameter $l$), using a{\color{black}n} out-of-sample period.  Figure \ref{fig:lag} depicts the MAE as a function of the number of lags and it is possible to see that MAE stops decreasing after $l=8$. Therefore, variables $W_{t-1}, ..., W_{t-8}$ are included in the final model. 

Additionally, the improvement obtained by including variables $CoT$, $GoD$, and ${}^{24H}_{}W^{}_{}$ was quantified. Table \ref{tab:FeaturesTest} summarizes the feature selection results. Each row is a model and each cell in the row is filled if the corresponding column (variable) is included in the model. The last two columns are the forecast performance obtained with the corresponding model. The performance metrics are negatively oriented, meaning small values are better. 

% Table generated by Excel2LaTeX from sheet 'tab'
\begin{table}[H]
	\centering
	\footnotesize
	\caption{Forecasting skill for different combination of features.}
	\begin{tabular}{c|ccc|c|c|c|c|c|c}
		\toprule
		\toprule
		\multicolumn{1}{c|}{\multirow{2}[4]{*}{\textbf{Model}}} & \multicolumn{3}{c|}{\textbf{Calendar}} & \multirow{2}[4]{*}{\textbf{Lags}} & \multirow{2}[4]{*}{\textbf{${}^{24H}_{}W^{}_{5,t}$}} & \multirow{2}[4]{*}{\textbf{CoT}} & \multirow{2}[4]{*}{\textbf{GorD}} & \multicolumn{2}{c}{\textbf{Error Metrics \color{black}{($m^3/h$)}}} \\
		\cline{2-4}\cline{9-10}    \multicolumn{1}{c}{} & \multicolumn{1}{|c}{\textbf{hour}} & \multicolumn{1}{c}{\textbf{wday}} & \textbf{month} &       &       &       &       & \textbf{MAE} & \textbf{CRPS} \\
		\hline
		M1    & \cellcolor[rgb]{ .906,  .902,  .902}x &       &       &       &       &       &       & 1428  & 1382 \\
		\cline{9-10}    M2    &       &       &       &       & \cellcolor[rgb]{ .906,  .902,  .902}x &       &       & 1077  & 1096 \\
		\cline{9-10}    M3    &   &       &       & \cellcolor[rgb]{ .906,  .902,  .902}x &       &       &       & 487   & 640 \\
		\cline{9-10}    M4    & \cellcolor[rgb]{ .906,  .902,  .902}x & \cellcolor[rgb]{ .906,  .902,  .902}x & \cellcolor[rgb]{ .906,  .902,  .902}x &       &       &       &       & 1153  & 1109 \\
		\cline{9-10}    M5    & \cellcolor[rgb]{ .906,  .902,  .902}x & \cellcolor[rgb]{ .906,  .902,  .902}x & \cellcolor[rgb]{ .906,  .902,  .902}x &       &       & \cellcolor[rgb]{ .906,  .902,  .902}x &       & 1118  & 1087 \\
		\cline{9-10}    M6    & \cellcolor[rgb]{ .906,  .902,  .902}x & \cellcolor[rgb]{ .906,  .902,  .902}x & \cellcolor[rgb]{ .906,  .902,  .902}x &       &       &       & \cellcolor[rgb]{ .906,  .902,  .902}x & 1117  & 1086 \\
		\cline{9-10}    M7    & \cellcolor[rgb]{ .906,  .902,  .902}x & \cellcolor[rgb]{ .906,  .902,  .902}x & \cellcolor[rgb]{ .906,  .902,  .902}x & \cellcolor[rgb]{ .906,  .902,  .902}x & \cellcolor[rgb]{ .906,  .902,  .902}x &       &       & 472   & 621 \\
		\cline{9-10}    M8    & \cellcolor[rgb]{ .906,  .902,  .902}x & \cellcolor[rgb]{ .906,  .902,  .902}x & \cellcolor[rgb]{ .906,  .902,  .902}x & \cellcolor[rgb]{ .906,  .902,  .902}x & \cellcolor[rgb]{ .906,  .902,  .902}x & \cellcolor[rgb]{ .906,  .902,  .902}x &       & 356   & 538 \\
		\cline{9-10}    M9    & \cellcolor[rgb]{ .906,  .902,  .902}x & \cellcolor[rgb]{ .906,  .902,  .902}x & \cellcolor[rgb]{ .906,  .902,  .902}x & \cellcolor[rgb]{ .906,  .902,  .902}x & \cellcolor[rgb]{ .906,  .902,  .902}x &       & \cellcolor[rgb]{ .906,  .902,  .902}x & 361   & 542 \\
		\cline{9-10}    M10   & \cellcolor[rgb]{ .906,  .902,  .902}x & \cellcolor[rgb]{ .906,  .902,  .902}x & \cellcolor[rgb]{ .906,  .902,  .902}x & \cellcolor[rgb]{ .906,  .902,  .902}x & \cellcolor[rgb]{ .906,  .902,  .902}x & \cellcolor[rgb]{ .906,  .902,  .902}x & \cellcolor[rgb]{ .906,  .902,  .902}x & 353   & 535 \\
		
		\bottomrule
		\bottomrule
	\end{tabular}%
	\label{tab:FeaturesTest}%
\end{table}%

The model conditioned to the hour of the day (M1) shows the worst performance with 1428 $m^3/h$ of MAE. Adding other calendar variables (M4), the error is improved with respect to M1 {\color{black}but}, when considering only the combination of calendar variables (M4) and $CoT$ or $GorD$ variable (M5 and M6), these models do not offer competitive performance in comparison with the inclusion of lagged information (M2 and M3). 

The combination of lags ($t-1 \ldots t-8$) and measurement from the previous day, results in a minor decrease of both metrics (model M7). Having $CoT$ or $GorD$ with lagged variables (M8 and M9) show similar errors $CRPS=356 m^3/h$ and $CRPS=361 m^3/h${\color{black},} respectively. {\color{black}However}, when both are added in M10, a further error reduction is attained with $CPRS=353 m^3/h$. The best model is achieved with all the features (M10).

\subsection{Performance Evaluation}

The forecasting skill was evaluated by splitting the intake dataset into an in-sample period, used for the initial parameter estimation and model selection, and an out-of-sample period, used to evaluate forecasting performance. 

The hyperparameters optimization was performed using Bayesian optimization \cite{Snoek2012} with 3-fold as cross validation on the in-sample period. Results are obtained for the out-of-sample period.

\begin{figure}[hbt!]
	\centerline{\includegraphics[trim=0cm 0.5cm 0cm 0cm, width=1\textwidth]{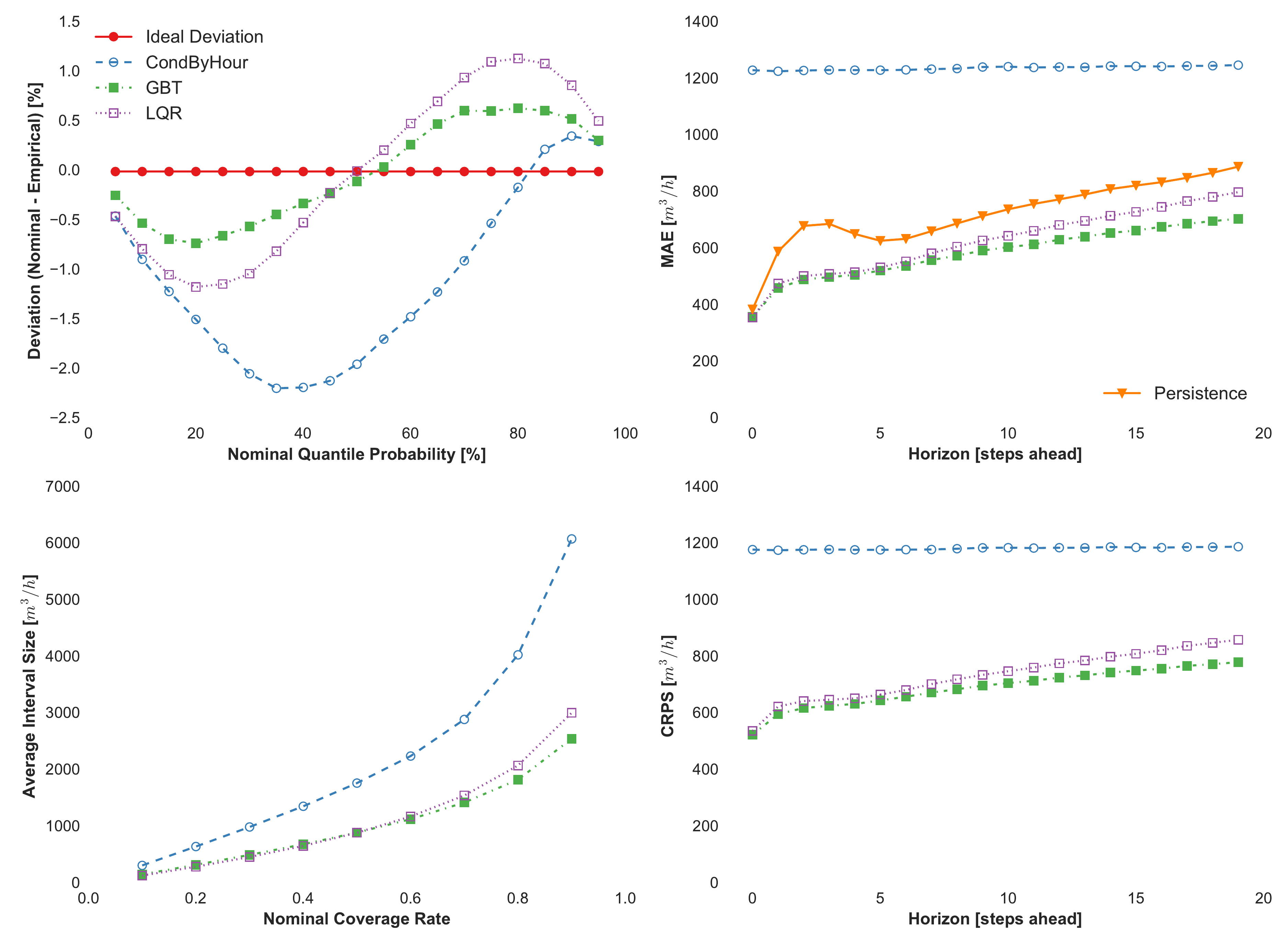}}
	\caption{Forecasting skill results for point and probabilistic forecasting. {\color{black}The best performance is obtained with the gradient boosting trees model.}}
	\label{fig:ww_results}
\end{figure}

The forecasting skill results are presented in Figure \ref{fig:ww_results} for all metrics.
The left-top plot depicts the difference from the “perfect calibration” (i.e., perfect match between nominal and empirical probabilities represented by the red line){\color{black},} averaged across all time horizons. The $CondByHour$ model showed the worst calibration of the three models, underestimating the forecast (negative deviation from perfect calibration). The remain{\color{black}ing} quantile models (GBT and LQR) underestimate quantiles below 50\% and overestimate quantiles {\color{black}up to} 50\%. In terms of calibration{\color{black},} GBT presents the best performance, being the deviation from the ideal (red line){\color{black},} smaller along all nominal quantile probabilities. Nevertheless, in all cases the deviation is within the interval -3\% and 3\%. 

Concerning sharpness (left bottom plot), the goal is to have small amplitude of forecast intervals for all coverage rates. Again, $CondByHour$ showed forecast intervals with wider amplitude,  average interval size varying from 327 to 6075 $m^3/h$ {\color{black}to increase} nominal coverage. $LQR$ and $GBT$ for the same nominal coverages showed much sharper intervals (almost half of the amplitude). 
Also{\color{black},} in Figure \ref{fig:ww_results}, the MAE and CRPS results are given for each step ahead and with the same units of the wastewater inflow $m^3/h$. We can see that $CondbyHour$ has the worst performance, with constant errors MAE$=1240 m^3/h$ and CRPS$=1185 m^3$ spread across the forecast horizon. Persistence results are only present for this metric, since it corresponds to point forecasts. As observed, persistence offers a very competitive forecasting skill. 
The GBT model (in green) presents the best performance in both metrics, following LQR in second place.

\setcounter{secnumdepth}{1}
\section{PPO Configuration}
\label{app:ppo}

\begin{table}[H]
	\centering
	\caption{Proximal policy optimization configuration}
	\begin{tabular}{l | c}
		\toprule
		\toprule
		variable & value \\
		\hline
		Num iterations    & 15k \\
		Policy layers & [64, 64] \\
		Value Function layers & [64, 64] \\
		Activation function & ReLU \\
		Timesteps per update & 750 \\
		Batch size & 250 \\
		Learning rate & 1.5e-4 (decaying exponentially) \\
		Entropy loss & 0.0 \\
		Clip param. & 0.2 \\
		$\gamma$ & 0.99 \\
		$\delta$ & 0.95 \\
		Reward scale & 600 \\
		Reward: $R^{+}$ & 3 \\
		Reward: $R^{-}$ & -600 \\
		Reward: $c_1$ & 1 \\
		Reward: $c_2$ & 0.4 \\
		\bottomrule
		\bottomrule
	\end{tabular}%
	\label{tab:ppo_config}%
\end{table}

\renewcommand\bibname{References}
\section*{References}
\bibliographystyle{model1-num-names}
\bibliography{Water_ref}

\begin{thebibliography}{46}
\expandafter\ifx\csname natexlab\endcsname\relax\def\natexlab#1{#1}\fi
\providecommand{\bibinfo}[2]{#2}
\ifx\xfnm\relax \def\xfnm[#1]{\unskip,\space#1}\fi
%Type = Techreport
\bibitem[{OEC(2009)}]{OECD2009}
\bibinfo{title}{Managing Water for All: An {OECD} Perspective on Pricing and
  Financing}, \bibinfo{type}{Technical Report}, OECD, \bibinfo{year}{2009}.
%Type = Article
\bibitem[{Guerrini et~al.(2015)Guerrini, Romano, Leardini, and
  Martini}]{Guerrini2015}
\bibinfo{author}{A.~Guerrini}, \bibinfo{author}{G.~Romano},
  \bibinfo{author}{C.~Leardini}, \bibinfo{author}{M.~Martini},
\newblock \bibinfo{title}{Measuring the efficiency of wastewater services
  through data envelopment analysis},
\newblock \bibinfo{journal}{Water Science and Technology} \bibinfo{volume}{71}
  (\bibinfo{year}{2015}) \bibinfo{pages}{1845--1851}.
%Type = Article
\bibitem[{Rasekh et~al.(2016)Rasekh, Hassanzadeh, Mulchandani, Modi, and
  Banks}]{Rasekh2016}
\bibinfo{author}{A.~Rasekh}, \bibinfo{author}{A.~Hassanzadeh},
  \bibinfo{author}{S.~Mulchandani}, \bibinfo{author}{S.~Modi},
  \bibinfo{author}{M.~K. Banks},
\newblock \bibinfo{title}{Smart water networks and cyber security},
\newblock \bibinfo{journal}{Journal of Water Resources Planning and Management}
  \bibinfo{volume}{142} (\bibinfo{year}{2016}) \bibinfo{pages}{1--3}.
%Type = Article
\bibitem[{Kebir et~al.(2014)Kebir, Demirci, Karaaslan, {\"U}nal, Dincer, and
  Aratc}]{Kebir2014}
\bibinfo{author}{F.~O. Kebir}, \bibinfo{author}{M.~Demirci},
  \bibinfo{author}{M.~Karaaslan}, \bibinfo{author}{E.~{\"U}nal},
  \bibinfo{author}{F.~Dincer}, \bibinfo{author}{H.~T. Aratc},
\newblock \bibinfo{title}{Smart grid on energy efficiency application for
  wastewater treatment},
\newblock \bibinfo{journal}{Environmental Progress \& Sustainable Energy}
  \bibinfo{volume}{33} (\bibinfo{year}{2014}) \bibinfo{pages}{556--563}.
%Type = Article
\bibitem[{Kalaiselvan et~al.(2016)Kalaiselvan, Subramaniam, Shanmugam, and
  Hanigovszki}]{Kalaiselvan2016}
\bibinfo{author}{A.~S.~V. Kalaiselvan}, \bibinfo{author}{U.~Subramaniam},
  \bibinfo{author}{P.~Shanmugam}, \bibinfo{author}{N.~Hanigovszki},
\newblock \bibinfo{title}{A comprehensive review on energy efficiency
  enhancement initiatives in centrifugal pumping system},
\newblock \bibinfo{journal}{Applied Energy} \bibinfo{volume}{181}
  (\bibinfo{year}{2016}) \bibinfo{pages}{495--513}.
%Type = Article
\bibitem[{Brion and Mays(1991)}]{Brion1991}
\bibinfo{author}{L.~M. Brion}, \bibinfo{author}{L.~W. Mays},
\newblock \bibinfo{title}{Methodology for optimal operation of pumping stations
  in water distribution systems},
\newblock \bibinfo{journal}{Journal of Hydraulic Engineering}
  \bibinfo{volume}{117} (\bibinfo{year}{1991}) \bibinfo{pages}{1551--1569}.
%Type = Article
\bibitem[{Menke et~al.(2016)Menke, Abraham, Parpas, and Stoianov}]{Menke2016a}
\bibinfo{author}{R.~Menke}, \bibinfo{author}{E.~Abraham},
  \bibinfo{author}{P.~Parpas}, \bibinfo{author}{I.~Stoianov},
\newblock \bibinfo{title}{Exploring optimal pump scheduling in water
  distribution networks with branch and bound methods},
\newblock \bibinfo{journal}{Water Resources Management} \bibinfo{volume}{30}
  (\bibinfo{year}{2016}) \bibinfo{pages}{5333--5349}.
%Type = Article
\bibitem[{Jacobusvan et~al.(2011)Jacobusvan, Zhang, and Xia}]{Jacobusvan2011}
\bibinfo{author}{A.~Jacobusvan}, \bibinfo{author}{J.~Zhang},
  \bibinfo{author}{X.~Xia},
\newblock \bibinfo{title}{A model predictive control strategy for load shifting
  in a water pumping scheme with maximum demand charges},
\newblock \bibinfo{journal}{Applied Energy} \bibinfo{volume}{88}
  (\bibinfo{year}{2011}) \bibinfo{pages}{4785--4794}.
%Type = Inproceedings
\bibitem[{L{\'o}pez-Ib{\'a}{\~n}ez et~al.(2005)L{\'o}pez-Ib{\'a}{\~n}ez,
  Prasad, and Paechter}]{Lopez-Ibaez2005}
\bibinfo{author}{M.~L{\'o}pez-Ib{\'a}{\~n}ez}, \bibinfo{author}{T.~D. Prasad},
  \bibinfo{author}{B.~Paechter},
\newblock \bibinfo{title}{Optimal pump scheduling: {R}epresentation and
  multiple objectives},
\newblock in: \bibinfo{booktitle}{Proceedings of the eighth International
  Conference on Computing and Control for the Water Industry},
  volume~\bibinfo{volume}{1}, pp. \bibinfo{pages}{117--122}.
%Type = Article
\bibitem[{Kernan et~al.(2017)Kernan, Liu, McLoone, and Fox}]{Kernan2017}
\bibinfo{author}{R.~Kernan}, \bibinfo{author}{X.~Liu},
  \bibinfo{author}{S.~McLoone}, \bibinfo{author}{B.~Fox},
\newblock \bibinfo{title}{Demand side management of an urban water supply using
  wholesale electricity price},
\newblock \bibinfo{journal}{Applied Energy} \bibinfo{volume}{189}
  (\bibinfo{year}{2017}) \bibinfo{pages}{395--402}.
%Type = Article
\bibitem[{Torregrossa and Capitanescu(2019)}]{Torregrossa2019}
\bibinfo{author}{D.~Torregrossa}, \bibinfo{author}{F.~Capitanescu},
\newblock \bibinfo{title}{Optimization models to save energy and enlarge the
  operational life of water pumping systems},
\newblock \bibinfo{journal}{Applied Energy} \bibinfo{volume}{213}
  (\bibinfo{year}{2019}) \bibinfo{pages}{89--98}.
%Type = Article
\bibitem[{Palensky and Dietrich(2011)}]{Palensky2011}
\bibinfo{author}{P.~Palensky}, \bibinfo{author}{D.~Dietrich},
\newblock \bibinfo{title}{Demand side management: {D}emand response,
  intelligent energy systems, and smart loads},
\newblock \bibinfo{journal}{IEEE Transactions on Industrial Informatics}
  \bibinfo{volume}{7} (\bibinfo{year}{2011}) \bibinfo{pages}{381--388}.
%Type = Article
\bibitem[{Bonvin et~al.(2017)Bonvin, Demassey, Pape, Maïzi, Mazauric, and
  Samperio}]{Bonvin2017}
\bibinfo{author}{G.~Bonvin}, \bibinfo{author}{S.~Demassey},
  \bibinfo{author}{C.~L. Pape}, \bibinfo{author}{N.~Maïzi},
  \bibinfo{author}{V.~Mazauric}, \bibinfo{author}{A.~Samperio},
\newblock \bibinfo{title}{A convex mathematical program for pump scheduling in
  a class of branched water networks},
\newblock \bibinfo{journal}{Applied Energy} \bibinfo{volume}{185}
  (\bibinfo{year}{2017}) \bibinfo{pages}{1702--1711}.
%Type = Article
\bibitem[{Zhuan and Xia(2013)}]{Zhuan2013}
\bibinfo{author}{X.~Zhuan}, \bibinfo{author}{X.~Xia},
\newblock \bibinfo{title}{Optimal operation scheduling of a pumping station
  with multiple pumps},
\newblock \bibinfo{journal}{Applied Energy} \bibinfo{volume}{104}
  (\bibinfo{year}{2013}) \bibinfo{pages}{250--257}.
%Type = Article
\bibitem[{Menke et~al.(2016)Menke, Abraham, Parpas, and Stoianov}]{Menke2016}
\bibinfo{author}{R.~Menke}, \bibinfo{author}{E.~Abraham},
  \bibinfo{author}{P.~Parpas}, \bibinfo{author}{I.~Stoianov},
\newblock \bibinfo{title}{Demonstrating demand response from water distribution
  system through pump scheduling},
\newblock \bibinfo{journal}{Applied Energy} \bibinfo{volume}{170}
  (\bibinfo{year}{2016}) \bibinfo{pages}{377--387}.
%Type = Article
\bibitem[{Afram et~al.(2017)Afram, Janabi-Sharifi, Fung, and
  Raahemifar}]{Afram2017}
\bibinfo{author}{A.~Afram}, \bibinfo{author}{F.~Janabi-Sharifi},
  \bibinfo{author}{A.~S. Fung}, \bibinfo{author}{K.~Raahemifar},
\newblock \bibinfo{title}{Artificial neural network ({ANN}) based model
  predictive control ({MPC}) and optimization of {HVAC} systems: {A} state of
  the art review and case study of a residential {HVAC} system},
\newblock \bibinfo{journal}{Energy and Buildings} \bibinfo{volume}{141}
  (\bibinfo{year}{2017}) \bibinfo{pages}{96--113}.
%Type = Inproceedings
\bibitem[{Jain et~al.(2018)Jain, Nghiem, Morari, and Mangharam}]{Jain2018}
\bibinfo{author}{A.~Jain}, \bibinfo{author}{T.~X. Nghiem},
  \bibinfo{author}{M.~Morari}, \bibinfo{author}{R.~Mangharam},
\newblock \bibinfo{title}{Learning and control using gaussian processes:
  towards bridging machine learning and controls for physical systems},
\newblock in: \bibinfo{booktitle}{ICCPS '18 Proceedings of the 9th ACM/IEEE
  International Conference on Cyber-Physical Systems}, \bibinfo{address}{Porto,
  Portugal}, pp. \bibinfo{pages}{140--149}.
%Type = Article
\bibitem[{Smarra et~al.(2018{\natexlab{a}})Smarra, Jain, de~Rubeis, Ambrosini,
  D'Innocenzo, and Mangharam}]{Smarra2018a}
\bibinfo{author}{F.~Smarra}, \bibinfo{author}{A.~Jain},
  \bibinfo{author}{T.~de~Rubeis}, \bibinfo{author}{D.~Ambrosini},
  \bibinfo{author}{A.~D'Innocenzo}, \bibinfo{author}{R.~Mangharam},
\newblock \bibinfo{title}{Data-driven model predictive control using random
  forests for building energy optimization and climate control},
\newblock \bibinfo{journal}{Applied Energy} \bibinfo{volume}{226}
  (\bibinfo{year}{2018}{\natexlab{a}}) \bibinfo{pages}{1252--1272}.
%Type = Article
\bibitem[{Smarra et~al.(2018{\natexlab{b}})Smarra, Jain, Mangharam, and
  D'Innocenzo}]{Smarra2018}
\bibinfo{author}{F.~Smarra}, \bibinfo{author}{A.~Jain},
  \bibinfo{author}{R.~Mangharam}, \bibinfo{author}{A.~D'Innocenzo},
\newblock \bibinfo{title}{Data-driven switched affine modeling for model
  predictive control},
\newblock \bibinfo{journal}{IFAC-PapersOnLine} \bibinfo{volume}{51}
  (\bibinfo{year}{2018}{\natexlab{b}}) \bibinfo{pages}{199--204}.
%Type = Misc
\bibitem[{Nybo et~al.(2014)Nybo, Kalles{\o}e, and Lauridsen}]{nybo2014method}
\bibinfo{author}{P.~J. Nybo}, \bibinfo{author}{C.~S. Kalles{\o}e},
  \bibinfo{author}{K.~G. Lauridsen}, \bibinfo{title}{Method for operating a
  wastewater pumping station}, \bibinfo{year}{2014}. \bibinfo{note}{U.S. Patent
  Application No. 14/133,938}.
%Type = Article
\bibitem[{Corominas et~al.(2018)Corominas, Garrido-Baserba, Villez, Olsson,
  Cort\'es, and Poch}]{Corominas2018}
\bibinfo{author}{L.~Corominas}, \bibinfo{author}{M.~Garrido-Baserba},
  \bibinfo{author}{K.~Villez}, \bibinfo{author}{G.~Olsson},
  \bibinfo{author}{U.~Cort\'es}, \bibinfo{author}{M.~Poch},
\newblock \bibinfo{title}{Transforming data into knowledge for improved
  wastewater treatment operation: {A} critical review of techniques},
\newblock \bibinfo{journal}{Environmental Modelling \& Software}
  \bibinfo{volume}{106} (\bibinfo{year}{2018}) \bibinfo{pages}{89--103}.
%Type = Article
\bibitem[{Torregrossa et~al.(2017)Torregrossa, Hansen, Hern\'andez-Sancho,
  Cornelissen, Schutz, and Leopold}]{Torregrossa2017}
\bibinfo{author}{D.~Torregrossa}, \bibinfo{author}{J.~Hansen},
  \bibinfo{author}{F.~Hern\'andez-Sancho}, \bibinfo{author}{A.~Cornelissen},
  \bibinfo{author}{G.~Schutz}, \bibinfo{author}{U.~Leopold},
\newblock \bibinfo{title}{A data-driven methodology to support pump performance
  analysis and energy efficiency optimization in waste water treatment plants},
\newblock \bibinfo{journal}{Applied Energy} \bibinfo{volume}{208}
  (\bibinfo{year}{2017}) \bibinfo{pages}{1430--1440}.
%Type = Article
\bibitem[{Fiter et~al.(2005)Fiter, G{\"u}ell, Comas, Colprim, Poch, and
  Rodr\'iguez-Roda}]{Fiter2005}
\bibinfo{author}{M.~Fiter}, \bibinfo{author}{D.~G{\"u}ell},
  \bibinfo{author}{J.~Comas}, \bibinfo{author}{J.~Colprim},
  \bibinfo{author}{M.~Poch}, \bibinfo{author}{I.~Rodr\'iguez-Roda},
\newblock \bibinfo{title}{Energy saving in a wastewater treatment process: an
  application of fuzzy logic control},
\newblock \bibinfo{journal}{Environmental Technology} \bibinfo{volume}{26}
  (\bibinfo{year}{2005}) \bibinfo{pages}{1263--70}.
%Type = Article
\bibitem[{Syafiie et~al.(2011)Syafiie, Tadeo, Martinez, and
  Alvarez}]{Syafiiea2011}
\bibinfo{author}{S.~Syafiie}, \bibinfo{author}{F.~Tadeo},
  \bibinfo{author}{E.~Martinez}, \bibinfo{author}{T.~Alvarez},
\newblock \bibinfo{title}{Model-free control based on reinforcement learning
  for a wastewater treatment problem},
\newblock \bibinfo{journal}{Applied Soft Computing} \bibinfo{volume}{11}
  (\bibinfo{year}{2011}) \bibinfo{pages}{73--82}.
%Type = Article
\bibitem[{del Olmo et~al.(2012)del Olmo, Gaudioso, and Nevado}]{del-Olmo2012a}
\bibinfo{author}{F.~H. del Olmo}, \bibinfo{author}{E.~Gaudioso},
  \bibinfo{author}{A.~Nevado},
\newblock \bibinfo{title}{Autonomous adaptive and active tuning up of the
  dissolved oxygen setpoint in a wastewater treatment plant using reinforcement
  learning},
\newblock \bibinfo{journal}{IEEE Transactions on Systems, Man, and Cybernetics,
  Part C (Applications and Reviews)} \bibinfo{volume}{42}
  (\bibinfo{year}{2012}) \bibinfo{pages}{768--774}.
%Type = Article
\bibitem[{Asadi et~al.(2016)Asadi, Verma, and Yang}]{Asadi2016}
\bibinfo{author}{A.~Asadi}, \bibinfo{author}{A.~Verma},
  \bibinfo{author}{K.~Yang},
\newblock \bibinfo{title}{Wastewater treatment aeration process optimization:
  {A} data mining approach},
\newblock \bibinfo{journal}{Journal of Environmental Management}
  \bibinfo{volume}{In Press} (\bibinfo{year}{2016}).
%Type = Article
\bibitem[{Han et~al.(2018)Han, Zhang, Liu, and Qiao}]{Han2018}
\bibinfo{author}{H.-G. Han}, \bibinfo{author}{L.~Zhang}, \bibinfo{author}{H.-X.
  Liu}, \bibinfo{author}{J.-F. Qiao},
\newblock \bibinfo{title}{Multiobjective design of fuzzy neural network
  controller forwastewater treatment process},
\newblock \bibinfo{journal}{Applied Soft Computing} \bibinfo{volume}{67}
  (\bibinfo{year}{2018}) \bibinfo{pages}{467--478}.
%Type = Article
\bibitem[{gui Han et~al.(2017)gui Han, Zhang, and fei Qiao}]{Han2017}
\bibinfo{author}{H.~gui Han}, \bibinfo{author}{L.~Zhang},
  \bibinfo{author}{J.~fei Qiao},
\newblock \bibinfo{title}{Data-based predictive control for wastewater
  treatment process},
\newblock \bibinfo{journal}{IEEE Access} \bibinfo{volume}{6}
  (\bibinfo{year}{2017}) \bibinfo{pages}{1498--1512}.
%Type = Article
\bibitem[{Wei and Kusiak(2015)}]{Wei2015}
\bibinfo{author}{X.~Wei}, \bibinfo{author}{A.~Kusiak},
\newblock \bibinfo{title}{Short-term prediction of influent flow in wastewater
  treatment plant},
\newblock \bibinfo{journal}{Stochastic Environmental Research and Risk
  Assessment} \bibinfo{volume}{29} (\bibinfo{year}{2015})
  \bibinfo{pages}{241--249}.
%Type = Article
\bibitem[{Zhang and Kusiak(2011)}]{Zhang2011}
\bibinfo{author}{Z.~Zhang}, \bibinfo{author}{A.~Kusiak},
\newblock \bibinfo{title}{Models for optimization of energy consumption of
  pumps in a wastewater processing plant},
\newblock \bibinfo{journal}{Journal of Energy Engineering}
  \bibinfo{volume}{137} (\bibinfo{year}{2011}) \bibinfo{pages}{159--168}.
%Type = Article
\bibitem[{Zhang et~al.(2012)Zhang, Zeng, and Kusiak}]{Zhang2012}
\bibinfo{author}{Z.~Zhang}, \bibinfo{author}{Y.~Zeng},
  \bibinfo{author}{A.~Kusiak},
\newblock \bibinfo{title}{Minimizing pump energy in a wastewater processing
  plant},
\newblock \bibinfo{journal}{Energy} \bibinfo{volume}{47} (\bibinfo{year}{2012})
  \bibinfo{pages}{505--514}.
%Type = Article
\bibitem[{Zeng et~al.(2016)Zeng, Zhang, Kusiak, Tang, and Wei}]{Zeng2016}
\bibinfo{author}{Y.~Zeng}, \bibinfo{author}{Z.~Zhang},
  \bibinfo{author}{A.~Kusiak}, \bibinfo{author}{F.~Tang},
  \bibinfo{author}{X.~Wei},
\newblock \bibinfo{title}{Optimizing wastewater pumping system with data-driven
  models and a greedy electromagnetism-like algorithm},
\newblock \bibinfo{journal}{Stochastic Environmental Research and Risk
  Assessment} \bibinfo{volume}{30} (\bibinfo{year}{2016})
  \bibinfo{pages}{1263--1275}.
%Type = Article
\bibitem[{Zhang et~al.(2016)Zhang, Kusiak, Zeng, and Wei}]{Zhang2016}
\bibinfo{author}{Z.~Zhang}, \bibinfo{author}{A.~Kusiak},
  \bibinfo{author}{Y.~Zeng}, \bibinfo{author}{X.~Wei},
\newblock \bibinfo{title}{Modeling and optimization of a wastewater pumping
  system with data-mining methods},
\newblock \bibinfo{journal}{Applied Energy} \bibinfo{volume}{164}
  (\bibinfo{year}{2016}) \bibinfo{pages}{303--311}.
%Type = Article
\bibitem[{Zhang et~al.(2015)Zhang, He, and Kusiak}]{Zhang2015}
\bibinfo{author}{Z.~Zhang}, \bibinfo{author}{X.~He},
  \bibinfo{author}{A.~Kusiak},
\newblock \bibinfo{title}{Data-driven minimization of pump operating and
  maintenance cost},
\newblock \bibinfo{journal}{Engineering Applications of Artificial
  Intelligence} \bibinfo{volume}{40} (\bibinfo{year}{2015})
  \bibinfo{pages}{37--46}.
%Type = Article
\bibitem[{Costanzo et~al.(2016)Costanzo, Iacovella, Ruelens, Leurs, and
  Claessens}]{Costanzo2016}
\bibinfo{author}{G.~Costanzo}, \bibinfo{author}{S.~Iacovella},
  \bibinfo{author}{F.~Ruelens}, \bibinfo{author}{T.~Leurs},
  \bibinfo{author}{B.~Claessens},
\newblock \bibinfo{title}{Experimental analysis of data-driven control for a
  building heating system},
\newblock \bibinfo{journal}{Sustainable Energy, Grids and Networks}
  \bibinfo{volume}{6} (\bibinfo{year}{2016}) \bibinfo{pages}{81--90}.
%Type = Inproceedings
\bibitem[{Lampe and Riedmiller(2014)}]{Lampe2014}
\bibinfo{author}{T.~Lampe}, \bibinfo{author}{M.~Riedmiller},
\newblock \bibinfo{title}{Approximate model-assisted neural fitted
  {Q}-iteration},
\newblock in: \bibinfo{booktitle}{2014 International Joint Conference on Neural
  Networks (IJCNN)}, \bibinfo{address}{Beijing, China}.
%Type = Article
\bibitem[{Yoo et~al.(2001)Yoo, Kim, Cho, Choi, and Lee}]{Yoo2001}
\bibinfo{author}{C.~K. Yoo}, \bibinfo{author}{D.~S. Kim},
  \bibinfo{author}{J.-H. Cho}, \bibinfo{author}{S.~W. Choi},
  \bibinfo{author}{I.-B. Lee},
\newblock \bibinfo{title}{Process system engineering in wastewater treatment
  process},
\newblock \bibinfo{journal}{Korean Journal of Chemical Engineering}
  \bibinfo{volume}{18} (\bibinfo{year}{2001}) \bibinfo{pages}{408--421}.
%Type = Article
\bibitem[{Koenker and Bassett(1978)}]{QR1978}
\bibinfo{author}{R.~Koenker}, \bibinfo{author}{G.~Bassett},
\newblock \bibinfo{title}{Regression quantiles},
\newblock \bibinfo{journal}{Econometrica} \bibinfo{volume}{46}
  (\bibinfo{year}{1978}) \bibinfo{pages}{33--50}.
%Type = Inproceedings
\bibitem[{Seabold and Perktold(2010)}]{statsmodels}
\bibinfo{author}{S.~Seabold}, \bibinfo{author}{J.~Perktold},
\newblock \bibinfo{title}{Statsmodels: Econometric and statistical modeling
  with python},
\newblock in: \bibinfo{booktitle}{Proceedings of the 9th Python in Science
  Conference}, volume~\bibinfo{volume}{57}.
%Type = Article
\bibitem[{Friedman(2001)}]{Friedman2001GBT}
\bibinfo{author}{J.~H. Friedman},
\newblock \bibinfo{title}{Greedy function approximation: A gradient boosting
  machine},
\newblock \bibinfo{journal}{Annals of Statistics} \bibinfo{volume}{29}
  (\bibinfo{year}{2001}) \bibinfo{pages}{1189--1232}.
%Type = Article
\bibitem[{Pedregosa et~al.(2011)Pedregosa, Varoquaux, Gramfort, Michel,
  Thirion, Grisel, Blondel, Prettenhofer, Weiss, Dubourg, Vanderplas, Passos,
  Cournapeau, Brucher, Perrot, and Duchesnay}]{Pedregosa2011}
\bibinfo{author}{F.~Pedregosa}, \bibinfo{author}{G.~Varoquaux},
  \bibinfo{author}{A.~Gramfort}, \bibinfo{author}{V.~Michel},
  \bibinfo{author}{B.~Thirion}, \bibinfo{author}{O.~Grisel},
  \bibinfo{author}{M.~Blondel}, \bibinfo{author}{P.~Prettenhofer},
  \bibinfo{author}{R.~Weiss}, \bibinfo{author}{V.~Dubourg},
  \bibinfo{author}{J.~Vanderplas}, \bibinfo{author}{A.~Passos},
  \bibinfo{author}{D.~Cournapeau}, \bibinfo{author}{M.~Brucher},
  \bibinfo{author}{M.~Perrot}, \bibinfo{author}{E.~Duchesnay},
\newblock \bibinfo{title}{Scikit-learn: Machine learning in python},
\newblock \bibinfo{journal}{Journal of Machine Learning Research}
  \bibinfo{volume}{12} (\bibinfo{year}{2011}) \bibinfo{pages}{2825--2830}.
%Type = Article
\bibitem[{Schulman et~al.(2017)Schulman, Wolski, Dhariwal, Radford, and
  Klimov}]{PPO_2017}
\bibinfo{author}{J.~Schulman}, \bibinfo{author}{F.~Wolski},
  \bibinfo{author}{P.~Dhariwal}, \bibinfo{author}{A.~Radford},
  \bibinfo{author}{O.~Klimov},
\newblock \bibinfo{title}{Proximal policy optimization algorithms},
\newblock \bibinfo{journal}{arXiv:1707.06347} \bibinfo{volume}{abs/1707.06347}
  (\bibinfo{year}{2017}).
%Type = Article
\bibitem[{Schulman et~al.(2015)Schulman, Moritz, Levine, Jordan, and
  Abbeel}]{GAE2015}
\bibinfo{author}{J.~Schulman}, \bibinfo{author}{P.~Moritz},
  \bibinfo{author}{S.~Levine}, \bibinfo{author}{M.~I. Jordan},
  \bibinfo{author}{P.~Abbeel},
\newblock \bibinfo{title}{High-dimensional continuous control using generalized
  advantage estimation},
\newblock \bibinfo{journal}{CoRR} \bibinfo{volume}{abs/1506.02438}
  (\bibinfo{year}{2015}).
%Type = Article
\bibitem[{Pinson et~al.(2007)Pinson, Nielsen, M{\o}ller, Madsen, and
  Kariniotakis}]{metricsPinson}
\bibinfo{author}{P.~Pinson}, \bibinfo{author}{H.~A. Nielsen},
  \bibinfo{author}{J.~K. M{\o}ller}, \bibinfo{author}{H.~Madsen},
  \bibinfo{author}{G.~N. Kariniotakis},
\newblock \bibinfo{title}{Non-parametric probabilistic forecasts of wind power:
  required properties and evaluation},
\newblock \bibinfo{journal}{Wind Energy} \bibinfo{volume}{10}
  (\bibinfo{year}{2007}) \bibinfo{pages}{497--516}.
%Type = Article
\bibitem[{Taieb et~al.(2016)Taieb, Huser, Hyndman, and Genton}]{Taieb2016}
\bibinfo{author}{S.~B. Taieb}, \bibinfo{author}{R.~Huser},
  \bibinfo{author}{R.~J. Hyndman}, \bibinfo{author}{M.~G. Genton},
\newblock \bibinfo{title}{Forecasting uncertainty in electricity smart meter
  data by boosting additive quantile regression},
\newblock \bibinfo{journal}{IEEE Transactions on Smart Grid}
  \bibinfo{volume}{7} (\bibinfo{year}{2016}) \bibinfo{pages}{2448--2455}.
%Type = Inproceedings
\bibitem[{Snoek et~al.(2012)Snoek, Larochelle, and Adams}]{Snoek2012}
\bibinfo{author}{J.~Snoek}, \bibinfo{author}{H.~Larochelle},
  \bibinfo{author}{R.~P. Adams},
\newblock \bibinfo{title}{Practical bayesian optimization of machine learning
  algorithms},
\newblock in: \bibinfo{booktitle}{Advances in Neural Information Processing
  Systems 25 (NIPS 2012)}, pp. \bibinfo{pages}{2951--2959}.

\end{thebibliography}

\end{document}